\documentclass[a4paper]{article}

\usepackage[pages=all, color=black, position={current page.south}, placement=bottom, scale=1, opacity=1, vshift=5mm]{background}
\SetBgContents{
	\tt This work is shared under a \href{https://creativecommons.org/licenses/by-sa/4.0/}{CC BY-SA 4.0 license} unless otherwise noted
}      

\usepackage[margin=1in]{geometry} 

\usepackage{amsmath}
\usepackage{amsthm}
\usepackage{amssymb}

\usepackage[utf8]{inputenc}
\usepackage{hyperref}
\hypersetup{
	unicode,
	pdfauthor={Author One, Author Two, Author Three},
	pdftitle={A simple article template},
	pdfsubject={A simple article template},
	pdfkeywords={article, template, simple},
	pdfproducer={LaTeX},
	pdfcreator={pdflatex}
}

\usepackage[sort&compress,numbers,square]{natbib}
\theoremstyle{plain}

\theoremstyle{definition}

\usepackage{graphicx, color}
\graphicspath{{fig/}}

\usepackage{algorithm, algpseudocode} 
\usepackage{mathrsfs} 

\usepackage{lipsum}
\usepackage[justification=centering]{caption}
\usepackage{rotating}
\usepackage[verbose]{placeins}

\title{Co-designing heterogeneous models:\\a distributed systems approach}
    \author{Marius-Constantin Ilau$^1$
    \and Tristan Caulfield$^1$
    \and 
    David Pym$^{1,2}$}

\date{
	$^1$University College London \\ \texttt{\{marius-constantin.ilau.18, t.caulfield, d.pym\}@ucl.ac.uk}\\%
	$^2$Institute of Philosophy, University of London  
    \\[2ex]%
	\today
}

\begin{document}
	\maketitle
    \begin{abstract}
	The nature of information security has been, and  probably will continue to be, marked by the asymmetric competition of attackers and defenders over the control of an uncertain environment. The reduction of this degree of uncertainty via an increase in understanding of that environment is a primary objective for both sides. Models are useful tools in this context because they provide a way to understand and experiment with their targets without the usual operational constraints. However, given the technological and social advancements of today, the object of modelling has increased in complexity. Such objects are no longer singular entities, but heterogeneous socio-technical systems interlinked to form large-scale ecosystems. Furthermore, the underlying components of a system might be based on very different epistemic assumptions and methodologies for construction and use. Naturally, consistent, rigorous reasoning about such systems is hard, but necessary for achieving both security and resilience. The goal of this paper is to present a modelling approach tailored for heterogeneous systems based on three elements: an inferentialist interpretation of what a model is, a distributed systems metaphor to structure that interpretation and a co-design cycle to describe the practical design and construction of the model. The underlying idea is that an open world interpretation, supported by a formal, yet generic abstraction facilitating knowledge translation and providing properties for structured reasoning and, used in practice according to the co-design cycle could lead to models that are more likely to achieve their pre-stated goals. We explore the suitability of this method in the context of three different security-oriented models: a physical data loss model, an organisational recovery under ransomware model and an surge capacity trauma unit model. \\
		
	\end{abstract}

	\section{Introduction}\label{sec:intro}

In the constantly evolving field of information security, models have proven to be extremely useful tools for understanding and managing the complexities of modern systems. They can be associated with a long list of benefits, including, but not limited to identification of vulnerabilities, prioritisation of security measures (for example, via systematic threat analysis, risk assessment, and attack simulation), design of effective security policies, better incident response and recovery planning, optimisation of resource allocation, improving stakeholder communication, and serving as vital artefacts for security awareness and training. Yet, modelling on its own is a complicated topic of inquiry, especially when considering heterogeneous systems. Our paper aims to provide a structured approach to designing, constructing, and reasoning about models of heterogeneous systems --- which are becoming increasingly necessary in security contexts. Additionally, it justifies the utility of the proposed methodology both abstractly, and explicitly via the presented model case studies.

All `good' models are alike, yet each `bad' model is `bad' in its own way; a reinterpretation of Tolstoy's famous introduction to Anna Karenina, serving to remind us that the ability to interpret phenomena around us and construct conceptual models based on such interpretations are subjective and strongly related to human cognition. Indeed, all `good' models are alike in the sense that they achieve a goal, whether that might be understanding, predicting, teaching, serving as a reference or guideline and so on. In other words, model quality is related to model instrumentality, but perhaps not directly to form. Each `bad' model is `bad' in its own way tells the other part of the story: just as interpretations are subjective, so are models, and although classes of `bad' models can be constructed based on form, this is significantly harder to do so based on interpretation. 

These are somewhat synthetic statements about what a `good' model is. We do not imply here that analytic statements exist, but that the above statements might simply change over time based on further observation. However, something we can clearly claim without a doubt is that the objects of modelling, or the phenomena and entities to be modelled have been increasing in complexity. The reasons are multiple: technological advances lead to larger and more interconnected systems on a global scale, social desiderata continuously bring changes to criteria that guide the construction and evaluation of systems and models in general and, new scientific paradigms generate different interpretations of phenomena which are then translated into new types of models. In a very Kuhnian sense, the paradigmatic shift cycle manifests itself not only at the level of modelling methodology but also at the level of interpretation choice, for each underlying phenomena to be included in a model. 

However, accepting this view brings forward some important implications towards modelling in the future. First of all, considering this continuous generative process of new theories and interpretations, we can assume that over time, the number and diversity of model epistemological assumptions will increase. Although this is not directly a negative effect, it raises the question of how could an increasing number of epistemically divergent models be integrated. A longer discussion about whether or not such models should be integrated is deferred to further work. Nevertheless, approaches such as multimethodology~\citep{mingers97, bowers2011} or scenario-based epistemic integration~\citep{bennett2006} represent attempts at solving this issue, at least at the level of conceptual models. 

Secondly, considering more procedural implications, the process of uncovering the relevant aspects of each epistemic assumption used in an aggregation of models is a process that requires the participation of multiple parties: modellers, experts, stakeholders, parties affected by it and so on. In this context, a perspective such as knowledge co-creation~\citep{regeer2009} or participatory design~\citep{spinuzzi2005} represent good starting points for constructing a process of knowledge translation between the interpretation of empirical realities and model truth. However, one must remember that models are not scientific theories, but a possibly instrumental step towards their development and because of that, their construction can be influenced by time and resource constraints. This can happen particularly when the goal of modelling is to provide a very practical solution to a developing crisis situation, as seen for instance during the Covid pandemic. 

The acceptance of these implications leaves the modeller facing the following question: how can models better describe the world and produce better results in the world if the underlying realities and systems become more complex and the number of parties involved in the modelling process and their assumptions about such realities increases and are not always in agreement? This is often the case for security models, where practical outcomes depend on the knowledge, goals, and acceptance of multiple involved participants.

In this paper, we propose an integrative methodology based on a mix of positivist and relativist desiderata, in an attempt to answer the above question from a practical perspective. Since the target of this modelling approach are heterogeneous systems, we argue that an inferentialist stance regarding the nature of models supported by a multimethodological view of model epistemology and ontology provides an open-enough interpretation for understanding such systems. Nonetheless, this plethora of interpretations needs to be managed somehow to produce practical results: we do so by translating the knowledge parts created by the participants in the modelling process, at the level of a translation zone, via the `distributed systems metaphor' and we take into account the different stages a model goes through during design and creation via the `co-design cycle'. 

In Section~\ref{sec:background}, we identify relevant methodological approaches to conceptual modelling grounded on philosophical positions such as referentialism, inferentialism and more pragmatic, engineering-focused perspectives. Section~\ref{sec:what_model} focuses on the authors' position regarding the nature of models, epistemology, ontology, and constructing a descriptive metric grounded in model inferentialism. In Section~\ref{sec:co_design}, we illustrate our co-design modelling methodology: we describe the principles of co-design, reason abstractly about their use in the area of modelling, identify caveats in the classical mathematical modelling cycle which can be addressed by co-design and then present our co-design modelling cycle. Section~\ref{sec:ds_met} contains further details regarding a specific area of the co-design cycle, namely the translation zone. We describe the trading zone and distributed systems metaphors, compare them at the level of benefits, constitutive elements and situations described and conclude they can be viewed as compatible. Then, we illustrate our notion of a translation zone explicitly by including elements of our distributed systems conceptualisation employed as a trading zone cultural tool. 

In Section~\ref{sec:case_studies}, we attempt to showcase the applicability of our method at the level of three different security oriented models --- a physical data loss model, an organisational ransomware recovery model and a trauma room surge capacity model --- constructed using the distributed systems metaphor and our co-design approach to modelling. We describe their goals, internal structure, representation choices and co-design approach employed. 

Lastly, Section~\ref{sec:conc} is used for concluding remarks and possible directions with further work. 


\section{Background}\label{sec:background}

Something currently regarded as common knowledge in the conceptual modelling literature is that any modelling theory must be able to provide an answer to the following two questions: what is the relationship between model and target and, how to produce accurate inferences about the target using the model. By target, we imply the heterogeneous system or ecosystem that the model is representing. Naturally, a third question can be raised: in which order should the two previous questions be asked and answered? This can be seen as a generative point for different philosophical positions upon which modelling research directions have developed. Based on this interaction between model and object of modelling, we can categorise approaches to modelling into two main accounts: referentialist and inferentialist. 

On one hand, referentialist views of modelling place utmost importance on defining the relationship between model and target from the very beginning. The relationship is a function mapping the structural elements of the model to the structural components of the target, and serves as justifiable explanation for the informational content of the model. Only after this initial step can the relationship be used as means of explaining inferences between model and target and evaluating their accuracy. Nonetheless, the above description can be seen as open to various interpretations, allowing for accounts differing at epistemic and ontological levels. For example, the direct referentialist positions of \citep{kripke1977, kripke1979}, \citep{kaplan1989, kaplan1989e, kaplan1990} or \citep{putnam1975} are theories of language claiming that the meaning of words and expressions lie in what they point out in the world. Because of that, a relationship between world and language must exist. They have been constructed as objective, realist focused reactions against the descriptivism and subjectivism of \citep{frege1892, frege1997}, which was proposing that the meanings of individual words are solely determined by their contribution to the thoughts conveyed within the sentences in which those words are used. Although different in some aspects, similar theoretical formulations of the relationship between model, or in the previous case language, and target can be found in \citep{suppes1966, sneed1976, vanfr1980, dacosta2003, lloyd2021}, focusing on notions of isomorphism, homomorphism or partial isomorphism. In rough terms, we can consider this direction as the classical, formal view of models.

However, conceptualising models as abstract formal structures is not the only direction that can be followed by proponents of referentialism. Accounts such as \citep{godfrey2009}, \citep{frigg2010, frigg2017, nguyen2022} or \citep{levy2015} form what is know as the fictions view of models. As noted by \citep{weisberg2012}, `proponents of the fictions accounts think that mathematical modeling involves engaging with fictions in a way that is analogous to reading stories or watching movies.' Depending on the epistemic view of fictions, the debate continues on whether models should be viewed as worlds of possibilities or products of the author's imagination. Although various differences can be observed at the level of metaphysical commitments, we consider them as referentialist because they further refine the relationship between model --- or remove the notion of model altogether and replace it with a process of idealization in the case of \citep{levy2015} --- and target. For a more detailed discussion, see \citep{weisberg2012}.

On the other hand, inferentialist approaches focus more on the empirical experimentation with phenomena as basis for model conceptualisation. Through such experimentation, an initial low level of detail model is constructed and further refined along a continuous process of inference. Only once a certain level of inference accuracy has been reached, the relationship between model and target can be determined. Different generally inferential perspectives include~\citep{suarez2004, contessa2007,suarez2010, dezon2009}, with~\citep{bueno2014, beisbart2012, beisbart2012m, kuorikoski2019} focusing specifically on simulation modelling. For a longer discussion about simulations, although not specifically focused on inferentialism, see~\citep{Durn2020}. 

Interestingly, most of these accounts have pronounced deflationist notes, in the sense that no deeper or unexplainable meaning is placed on the model-target relationship. As Kuorikoski puts it, `We seem to learn something genuinely new about the world by manipulating an artificial surrogate system and then “observing” what the end result is.'~\citep{kuorikoski2019}, but the seeming aspect here is relevant. The model is not viewed as producing a similar type of knowledge as empirical experimentation with phenomena, but rather as a reasoning helping tool whose epistemic value should be analysed in terms of the scope and reliability of inferences. This is achieved by keeping record and managing the doxastic commitments and, inherently, the assumptions and bias introduced by the modellers explicitly.

Nevertheless, not all modelling approaches can be so easily classified, especially given the fact that some can be rooted in a more engineering focused direction. A good example here is \citep{thalheim2010}, which attempts to form a theoretical basis for conceptual modelling concentrating on engineering and design principles. Without a clear adherence to either inferentialist or referentialist directions, the author describes the act of modelling as a tripartite activity decomposed into modelling language constructs, application domain gathering and engineering. Because of the importance shown in the interaction with the application domain and, in the definition of model properties such as utility, we tend to believe the author describes an inferentialist practice of model construction. The general implication seems to be that a modeller must first interact with an application --- in this case a database --- to be able to construct a conceptual schema of it. However, the methods for modelling language constructs seem to imply a concept of denotation closer to~\citep{frigg2016f}, even if used over empirical realities --- if a software artefact is considered one --- not products of imagination. When describing necessary properties for models in the future,~\citep{thalheim2010} includes the ability to perform an adjustable selection of principles depending on modelling goals, model suites with explicit model association, the explicit treatment of model value and adequate representation variants of models. In the further sections, we shall attempt to answer to some of the questions that arise from these requirements. Nonetheless, we acknowledge that the modelling suites and the co-design methodology presented in~\citep{schewe2019, schewe2019co} and exemplified in~\citep{schewe2005} are attempts worth undertaking in understanding and managing model heterogeneity. A similar approach can be seen in the work of~\citep{demjaha2021, demjaha2023}, with a particular focus on the human oriented factors involved in modelling as a collective, rather than singular activity.

With that in mind, our main argument follows this direction: since modelling is a collective activity involving parties with various beliefs, goals and interpretation criteria attempting to produce a singular representation of a heterogeneous target, the question to ask should not be what is the best representation of the target, but how to manage the underlying aspects, the so called unstated assumptions, that so often end up deviating the co-created model from a more favourable manifested configuration. That is to say, multiple methodological iterations of the modelling process will lead to an improve in model quality only if they do not end up propagating a misalignment of goals and interpretation introduced by the involved parties.

\section{What is a model?}\label{sec:what_model}

In the previous sections, we have attempted to briefly describe the theoretical context into which our research could be placed in. In that context, we have identified a series of elements that we wish to further develop in our modelling theory: managing model heterogeneity, explicitly accounting for the model goal and conceptualising modelling as a collective activity. However to be able to describe, relate to, and employ such concepts, we first require a basis for what our notion of model means and, more importantly, implies. Therefore, in Subsection~\ref{subsec:model_def}, we define what a model is and describe the implications of the definition with regards to model epistemology and ontology with the help of two examples. In Subsection~\ref{subsec:model_qual} we describe three relevant model qualities --- namely conceptuality, formality and executability --- together with a list of benefits that can justify their use beyond simple categorization.

\subsection{Model definition}\label{subsec:model_def}

We start with our definition of model: A model is a representation of a target serving a purpose. The choice of a simple, yet generic definition here is purposeful, to emphasise the constituent elements. 

First of all a model is a representation of a target. That is, in the words of \citep{burkhardt1991}, a mental object standing for something else. Naturally, various theories of representation have been developed and used underneath both the inferentialist and referentialist approaches. To some extent, our position on the matter can be considered as deflationary because of two reasons: we do not impose a series of necessary and sufficient criteria for what representation is and we do believe that representation in a specific domain is deeply connected to the practices of that domain. In the formulation by Su\'arez, `it is impossible, on a deflationary account, for the concept of representation in any area of science to be at variance with the norms that govern representational practice in that area … representation in that area, if anything at all, is nothing but that practice'~\citep{suarez2015}. Although deflationary, we lean towards a cognitivist perspective similar to \citep{oltramari2005} in the sense that we adhere to the separation of presentation and representation --- `the notion of presentation deals with the dynamic forms of cognition while representation is mainly bound to an abstraction level'. 

Furthermore, the nature and complexity of the target are relevant. On a theoretical level, that is because the interpretation of the presentation of the target influences the decision of how to represent it. Model targets can be both empirical and non-empirical, vary in complexity from singular phenomena to ecosystems and accumulative tendencies exist in both directions. Empirically, people construct more complex systems over time to fulfill their needs --- this can be seen, for example, in a review of the development of telecommunication systems~\citep{hugill1999} or other similar historical presentations. Non-empirically, both scientific theories and artistic and literary currents go through accumulative and revolutionary stages in a Kuhnian sense~\citep{kuhn2012}, except the criteria for what starts the transition between phases is different. From a modelling perspective, this set of criteria is not relevant. The focus should remain on a method of model design and construction which takes into account this increasing variety of model targets. For the specific cases of modelling without a specific target as described by \citep{weisberg2012} --- generalized modeling, hypothetical modeling, and targetless modeling --- we simply view the target as highly conceptual.

Secondly, a model serves a purpose. While this is generally accepted in the literature, and works such as \citep{forrester97}, \citep{dhrymes72} and \citep{radzicki88} explicitly describe the importance of the model goal at the level of validation, they implicitly assume that such a goal is singular at a given time in the life-cycle of the model. Particularly, Forrester claims that `the validity of a model should not be separated from the validity and the feasibility of its goal'~\citep{forrester97} and Dhrymes concludes that `validation becomes a problem-dependent or decision-dependent process, differing from case to case as the proposed use of the model under consideration changes'~\citep{dhrymes72}. However, our position on the influence of the modelling goal on the actual model is more extensive and does not include only the validation aspects. Following the presentation and representation distinction from above, we argue that the modelling goal influences both. At the level of presentation, although the goal cannot affect the empirical manifestation of phenomena, it can surely impact their interpretation. This usually translates into a limitation of decisions with regards to the model representation simply because some aspects are left out. We briefly illustrate these influences by looking at two models targeting the same phenomenon --- namely the decline of the population of scallops in the French Bay of St Brieuc --- but with differently stated goals. The models we look at are \citep{callon1984} and \citep{fresard2006}. The first one is highly conceptual and attempts to determine possible reasons for the decline in the scallop population, whereas the second one is mathematical and simulational, and determines generalised equilibrium manifestations for the scallop population under different competition and environmental assumptions. 

The first model uses a specific framework, namely the sociology of translation, to represent the actors of this environmental ecosystem --- fishermen, scallops, predators, scientists, local communities --- as going through a set of phases --- problematisation, interassessment, enrollment and mobilisation --- which lead to the development of social relations between them. We do not question here the possibility for the development of a social relationship between a scallop and a scientist. During this process of representation, the scientists end up as representatives for all the other actors and misrepresent their characteristics, this leading to controversies which in the end destroy the constructed actor-network. Therefore, the model does not produce explicit results, but rather a process to be observed, which could lead to them. 

In the case of the second model, the situation is different because the input is mostly numerical data, manipulated by the model through statistical operations and then interpreted at the end. Briefly, simulations are used to show that controlling invasive species can lead to an increase in the scallop population, and furthermore to quantifiable economic benefits. 

In the end, both models manage to reach their stated goals. The first one inherently shows how a misrepresentation problem can lead to a lack of practical solutions to the population issue and in the end to even more population decline whereas the second one produces its desired general solution. The chosen interpretation for the phenomenon is clearly different in the two cases and does lead to different model representations and types of solutions, even if considering the same target. Nevertheless, constructing models in an ontological and epistemic silo is not the only possibility. If heterogeneous systems are scientifically agreed upon, we believe models are also heterogeneous, in the sense that they should be able to construct inter-operable representations of phenomena with wildly different epistemic and ontological bases. Because of that, we argue that a general notion of model should be viewed as `multimethodological'~\citep{mingers97} epistemically and ontologically. Similarly to a distributed system case, the management of this differences allows the model to manifest coherently. Section~\ref{sec:ds_met} further details this matter. 

\subsection{Model qualities}\label{subsec:model_qual}
Having described our notion of model, the modelling goal, the implications of the interpretation of presentation on representation and inherently on the epistemic and ontological aspects of a model, we now focus on how to characterize a model. Since our conceptualization of models is driven by a pragmatic commitment to practicality, we cannot base our characterization solely on presentation or interpretation. Therefore, we have chosen the manifestation of representation, or the means of construction because of its direct relationship with language constructs and their determinable nature at a specific point in time.

Based on its construction and manifestation, a model has 3 main qualities: conceptuality, executability and formality. In~\citep{caulfield2021meta}, we present them under the name of `Triangle Framework' and empirically explore their appropriateness for describing the components of models, in a cyber-security context. 

\begin{itemize}
    \item[-]\textit{Conceptuality} ---  The conceptualization of a model pertains to how its fundamental components and the connections between them are present and explicitly conveyed through precise natural language, pictures or diagrams, for instance. To some extent, this can be viewed as the size of the directly expressed ontology.
    
    \item[-]\textit{Formality} --- This refers to the degree to which the elements and relationships of a model are expressed using formal constructs. For example, models might be expressed as systems of equations or logical formulae. 
    
    \item[-]\textit{Executability} --- This represents the degree to which the elements and relationships of a model manifest themselves in a physical or computational environment through series of iterative steps. 
\end{itemize}

Various model categorization attempts have been undertaken in previous years with their focus on the so called most basic constitutive elements. An example can be seen in~\citep{weisberg2012}, where concrete, mathematical and computational models are clearly separated. The main difference between such an approach and ours is that in our view a model can manifest degrees of this qualities at the same time. For example, a simulation model may be comprised of two different sets of simulations: one based on a system of equations, and another based on a state machine which was constructed ad-hoc from empirical observations. After the simulations are completed, the results might be presented visually through graphs or formulated in natural language. In such a case, we do not believe that labelling the model as computational is enough because each decision to use a certain construction technique is related to the model goal. Furthermore, such an approach would lead to viewing most presentations of a model --- interestingly, the constructed representation of a model ends up becoming a surrogate presentation for the model target --- as conceptual and therefore omitting its previous configuration and implications from analysis. Not taking into account all these decisions can lead to model misrepresentation and misuse. 

As previously said, these qualities coexist and models may have components that exhibit characteristics of all three. Additionally, they trade off against one another; that is, a highly conceptual model will be less formal and executable, or a highly executable will be less conceptual and formal. This happens because the process of designing and constructing a model has resource constraints and favouring one of the three qualities based on the context of use can imply a higher chance of achieving the model goal. For example, using a Bayesian model in a court of law will most likely not have a high chance of success --- for details see~\citep{dawid2002bayes} --- but a conceptual model based on argumentation with the exact same conclusions might do it. More malicious examples could be constructed here, because favouring a specific subset of the elements of the presentation of a phenomena combined with a construction technique that is considered appealing by a target audience is exactly how manipulation techniques work. We will not further pursue this direction, but we clearly state that the difference between manipulation and focusing on the model goal while considering some elements irrelevant for the context exists, and lays with the parties involved in the modelling process. In the case of manipulation, some elements are known to be relevant but specifically omitted so that the final results would be different. In the other case, the elements are omitted precisely because they would not impact the final results.

\begin{figure}[hbt!]
    \centering
    \includegraphics[scale=0.3]{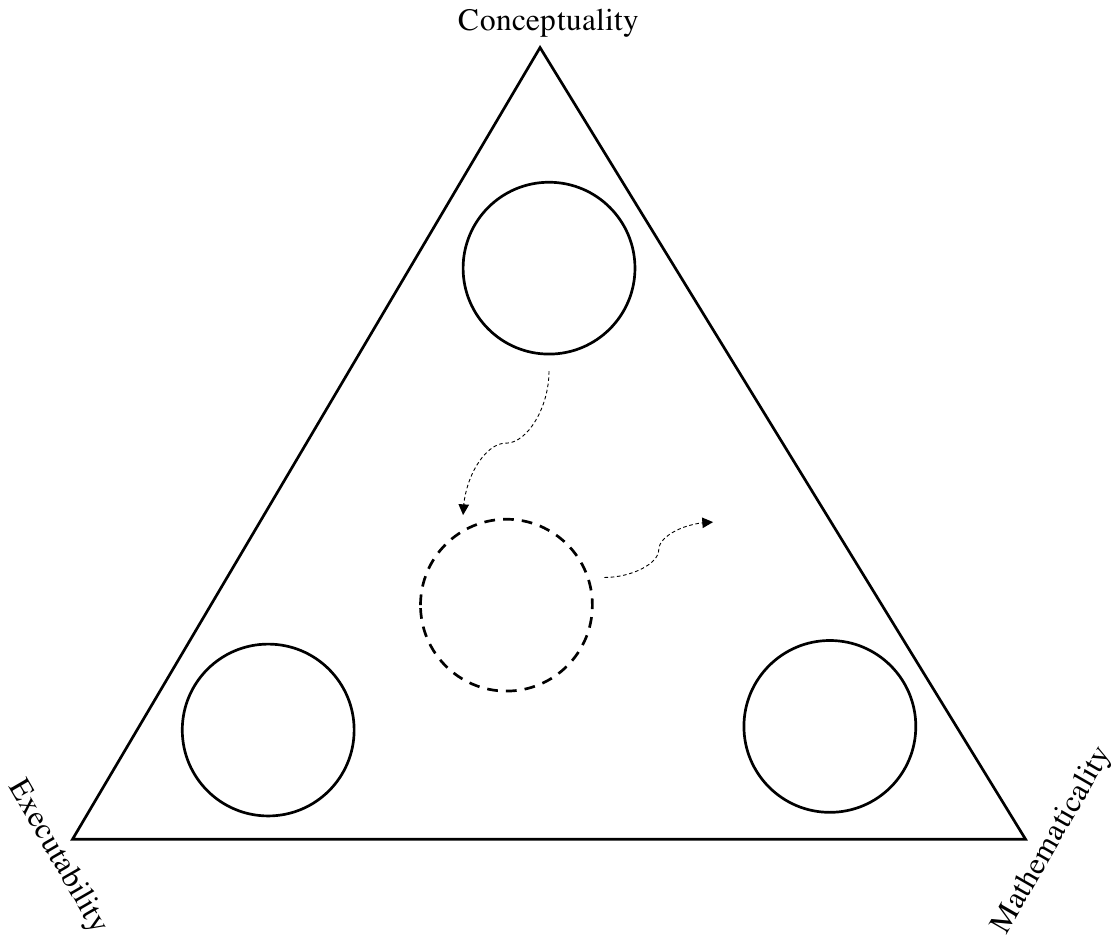}
    \caption{The triangle framework~\citep{caulfield2021meta}}
    \captionsetup{justification=centering}
    \label{fig:triangle}
\end{figure}

While we discuss the degree of model qualities, it's important to note that we don't imply a precise measure --- there are no units for formality, executability, or conceptuality, and their quantification is subjective. Nevertheless, these qualities, along with the triangular framework in general, provide us with a language to structure and discuss models. Figure~\ref{fig:triangle} illustrates how we think of these three types of model and their relationships: in a given model, the relative significance of each of the components determines, by proximity, the position of the model within the triangle; furthermore, the position of the model may change as it evolves during its construction. 

\FloatBarrier

We suggest that the importance of our proposed framework goes beyond its use for 
categorizing, and hence understanding the relationships between, extant models 
and types of models. Specifically, we suggest that it can: 
\begin{itemize}
    \item[-] Assist the parties involved in modelling in determining which property to emphasize to enhance the likelihood of more accurately representing the target, in alignment with the modeling goals;  
    \item[-] Minimize the likelihood of creating a model that is impractical for achieving the modeling goals; 
    \item[-] Serve as a common base for argumentation, and therefore language for both structuring the process of design and construction and, formulation of requirements;
    \item[-] Offer a way of analysing a model through all the design and construction stages rather than just at the end 
    \item[-] Reduce development time because a direction for model evolution can be determined and decided upon, rather than just emerge during construction;
    \item[-] Lead to a way of comparing models based on the properties of sub-models. 
\end{itemize}









\section{The co-design modelling methodology}\label{sec:co_design}

In this section, we focus on constructing a methodological approach to modelling based on updating the classical, mathematical modelling cycle by including ideas derived from the principles of co-design. In Subsection~\ref{subsec:co_design} we explicitly describe what we mean by co-design. In Subsection~\ref{subsec:classical_cycle}, we briefly analyse the classical mathematical modelling cycle and showcase some of its caveats that need addressing. Lastly, in Subsection~\ref{subsec:co_design_cycle}, we present our version of co-design cycle.

\subsection{What is co-design?}\label{subsec:co_design}

As defined in~\citep{kleinsmann2008}, `co-design is the process in which actors from different disciplines share their knowledge about both the design process and the design content. They do that in order to create shared understanding on both aspects, to be able to integrate and explore their knowledge and to achieve the larger common objective: the new product to be designed'. This formulation focused on the notion of process was developed as a response to the need of structuring and addressing collaboration and communication problems --- in areas such as information gathering and sharing, problem analysing and understanding, concept generation and adoption, conflict resolution and others --- observed in the design literature at the level of multidisciplinary teams. For a thorough analysis of the topic, see~\citep{bucciarelli1988, bucciarelli1994, cross95, badke1999, bucciarelli2002, badke2007}.

For the purpose of clarity, we make the distinction between co-creation and co-design. As defined in~\citep{sanders2008}, co-creation represents `any act of collective creativity, i.e. creativity that is shared by two or more people' and is a more general concept than co-design because it refers to creative collaboration throughout more than the design oriented stages in the life-cycle of a product. From a modelling perspective, although co-creation could be viewed as a desiderata~\citep{voinov2016}, usually it might be simply impractical to involve stakeholders, domain experts and users in all the model construction phases. Typical reasons for this impracticality are related to either resource constraints --- for example personal time constraints or remuneration issues, because modelling is not always part of stakeholders' or domain experts' workload --- or a lack of understanding of the practical construction methods, especially in the case of formal or executable models. Nevertheless, since model design stages heavily lean towards conceptuality as explained in Section~\ref{subsec:model_qual}, the above issue of lack of understanding becomes more addressable. Because of this pragmatic reason, we choose to focus on co-design rather co-creation.

However, the idea of a co-design focused modelling process is not entirely new. Approaches such as user-centered design~\citep{abras2004} or participatory design~\citep{spinuzzi2005} have led to the appearance of participatory modelling~\citep{basco2017}, which represents an attempt at applying participatory design principles at the level of modelling. As defined in~\citep{voinov2018} participatory modelling is a `purposeful learning process for action that engages the implicit and explicit knowledge of stakeholders to create formalized and shared representations of reality'. Authors such as~\citep{alam2002, kujala2003, muller2012} associate a plethora of advantages such as a higher quality of system requirements, higher system quality, a better fit
between the system and users’ needs, improved satisfaction and mutual understanding of users or customers, development of differentiated new services with unique benefits, reduced development time, education of users, enhancing communication and cooperation
between different people, and joint creation of new ideas, to the use of such collaborative design methodologies, particularly when related to service design and development.

In the context of this paper, given our deflationist views, we argue that the representations constructed through our modelling process need not necessarily be fully formalised, but rather discussed, justified and clearly agreed upon during the modelling process. Still, in Section~\ref{sec:ds_met}, we describe some of the requirements and benefits of employing a formalised distributed systems metaphor for structuring the co-design aspects of model design. We consider this step as process optimisation rather than baseline necessity, particularly because methods such as SSM (Soft Systems Methodology)~\citep{checkland2020} have been successfully used in the collaborative creation of conceptual models without formalisation. Even though a formal description of SSM has been achieved in~\citep{sagoo1998} via Petri Nets, this has not led to a significant transformation of the process of constructing conceptual models of systems using SSM. 

\subsection{The classical modelling cycle}\label{subsec:classical_cycle}

Having previously focused on co-design and its advantages, we now turn our attention to the actual process of designing and constructing models in the context of heterogeneous systems. 

Although from a historical perspective, conceptual models have existed since at least the early days of humanity, a generally agreed upon process for their design and construction still remains an active topic of research today. However, this is not the case for formal or mathematical models, due to their usually positivist epistemic tendencies. Thus, the process of model design and construction in the case of formal models can be considered in a more mature state than its conceptual counterpart and therefore, more properties of models constructed in such a way are known. Because of this, we have chosen to construct our integrative methodology starting from this classical, mathematical modelling cycle and we are attempting to expand it for the more complex case of heterogeneous systems. We acknowledge the fact that the process could be reversed, perhaps by starting from a methodology less focused on formality, but we argue that the set of properties provided by the classical mathematical modelling cycle, if underpinned by a `good' metaphor, reasoning framework and tools for practical implementation are significantly harder to produce if starting for example from SSM~\citep{checkland2020}. More about this in Section~\ref{sec:ds_met}.

\begin{figure*}[t]
     \centering
     \includegraphics[scale=0.4]{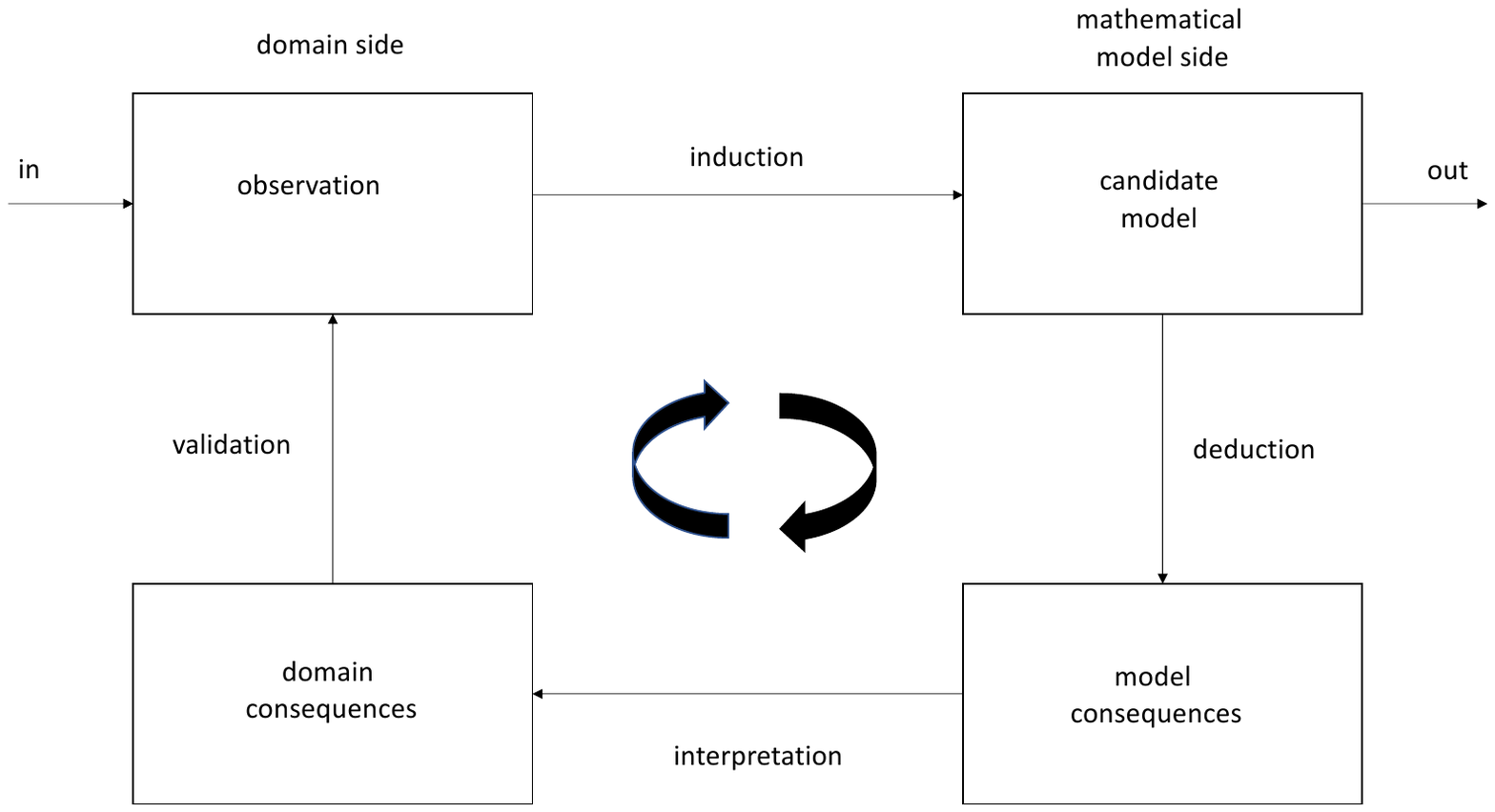}
     \caption{The classical mathematical modelling cycle \citep{caulfield2021meta}}
     \label{fig:math-cycle}
 \end{figure*}

Figure~\ref{fig:math-cycle} details what we mean by the classical mathematical modelling cycle. As previously described in~\citep{caulfield2021meta, demjaha2021} this is a four step iterative process focusing on:

\begin{itemize}
        \item[-] observing the manifestation of a phenomenon in its domain, 
        \item[-] constructing a candidate model based on the observations via induction, 
        \item[-] deducing the mathematical consequences of the model, 
        \item[-] interpreting the consequences of the model in the domain, 
        \item[-] validating the model by comparing the domain interpreted consequences with the initial observations
    \end{itemize}

Furthermore, this process' is underpinned by a set of usually unstated assumptions, such as: 

\begin{itemize}
        \item[-] The structure and behaviour of the domain is clearly understood in conceptual or engineering terms.    
        \item[-] The data that can be collected about the domain is essentially unambiguously identified. 
        \item[-] The questions that the model is intended to address are identified independently of the detailed design choices required for the construction of a model.  
\end{itemize}

However, given the nature of heterogeneous systems, this is hardly the case. For example, the structure and behaviour of the domain are harder to grasp because the domain is composed of multiple sub-domains with different structures, behaviours, incentives and so on. Naturally, a clear understanding of such domains requires access to knowledge from different scientific disciplines, if such knowledge already exists. Furthermore, this issue extends to the domain data collection: multiple data types might be available, and the process of unambiguously identifying the relevant aspects from it can be more complex, especially when cross-checking between different types. This can be particularly difficult if the data has already been recorded before that start of the modelling process, since relevant features might be missing from quantitative data and additional clarifications might be required for qualitative data. Last but not least, independently identifying and fixating the questions the model should address before knowing the available resources --- data, expertise, even deployment environment --- might not always be the best idea. We do not attempt to diminish the importance of setting a clearly defined goal from the very beginning here. However, goals can be attained in different ways: for example, if the goal of a model is to stop ransomware from infecting a computer in a specific organisation, one could ask how does ransomware behave in general, or what are the top three indicators that an email attachment contains ransomware. The distinction here is clear: although the first question is not incorrect, it shows a lack of information regarding the environment in which the model will be deployed. If the only available entry point is via email, there is no need to represent possible infection via USB ports. In other words, even if the overall model goal is pre-stated, the model scope and, inevitably the exact questions to be answered should co-evolve with the model. As described by~\citep{steen2013}, `in design thinking, problems and possible solutions are explored and developed and evaluated simultaneously in an iterative process: A “design process involves finding as well as solving problems” so that “problem and solution co-evolve.” Design thinking is needed to cope with “wicked problems”—
problems that cannot be clearly defined using “facts” at the start of a project and that cannot be solved by selecting a “best” solution'. 


\subsection{The co-design cycle}\label{subsec:co_design_cycle}

Building on the previous subsection, we can clearly state our starting point --- the classical mathematical modelling cycle --- and our desired outcome: a modelling cycle that would explicitly acknowledge and facilitate the understanding of both the structure and behavior of the domain and phenomenon under study, the unambiguous identification of relevant entities and relationships from the available data and the co-evolution of model and scope while conserving the pre-stated goal. 

\begin{sidewaysfigure}[hbt!]
    \centering
    \includegraphics[scale=0.7]{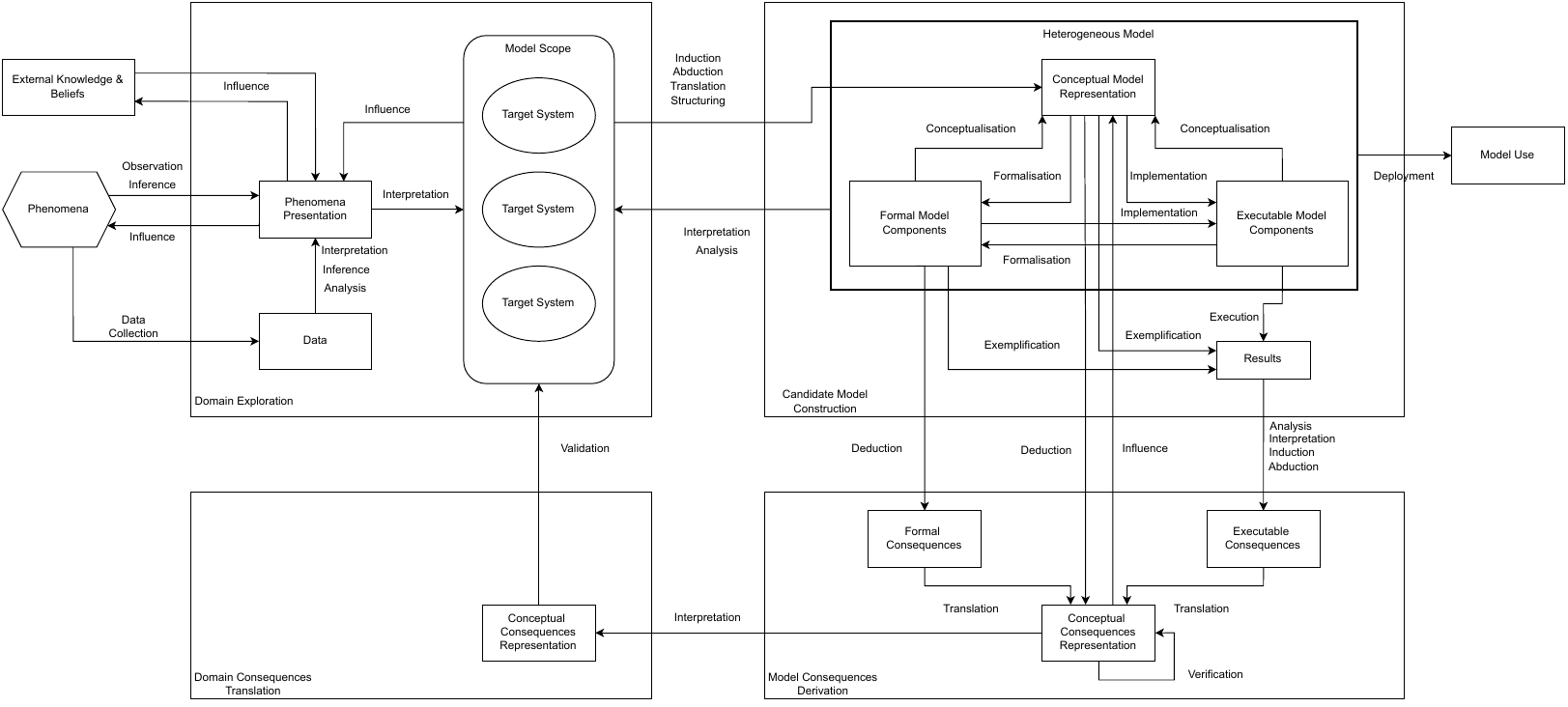}
    \caption{The co-design cycle}
    \captionsetup{justification=centering}
    \label{fig:cycle}
\end{sidewaysfigure}

The authors' previous work in~\citep{caulfield2021meta, demjaha2021} represents preliminary attempts at constructing a modelling cycle based on the principles of co-design, which can be observed in Figure~\ref{fig:sp_co_cycle}. In the following paragraphs, we further develop this cycle by explicitly accounting for the definitions in Subsection~\ref{subsec:model_def} and integrating it with the qualities described in Subsection~\ref{subsec:model_qual}. The resulting co-design cycle can be observed in Figure~\ref{fig:cycle}. We further describe the components:

\begin{figure}[hbt!]
    \centering
    \includegraphics[scale=0.4]{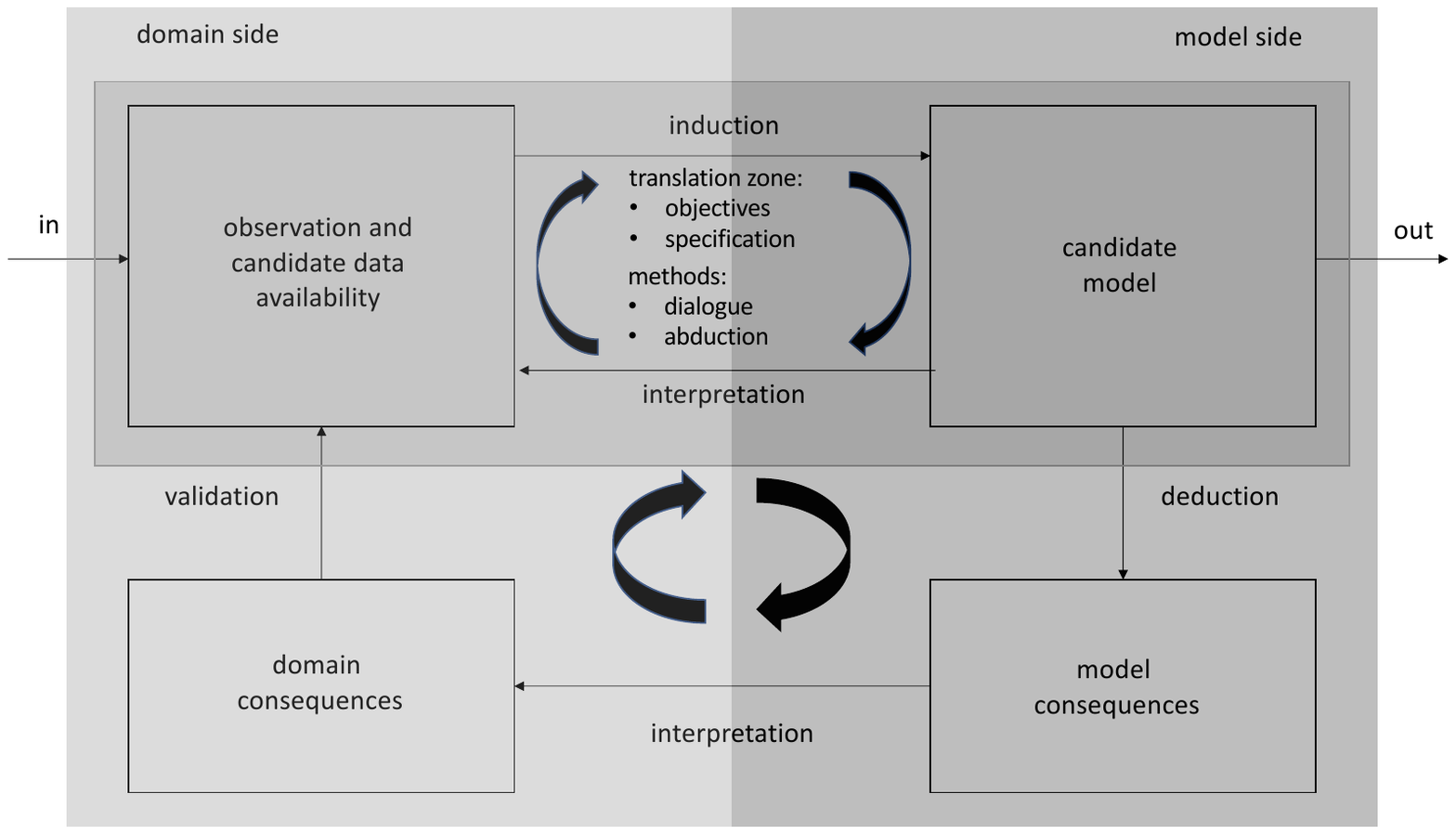}
    \caption{The placement of the translation zone on the simplified co-design cycle~\citep{caulfield2021meta}}
    \captionsetup{justification=centering}
    \label{fig:sp_co_cycle}
\end{figure}
\FloatBarrier 

\subsubsection{Domain exploration}

The primary objective of the domain exploration phase is represented by the construction of an initial model scope. Since the model scope is constituted by the list of target systems, their relevance and representation quality criteria, this implies the participants posses or have acquired a degree of understanding of the phenomena under study. Each participant constructs an initial conceptualisation of the phenomena --- the phenomena presentation ---  based on direct observation and inference, prior knowledge, beliefs \& interpretation, and analysis and interpretation of available data.  By interpreting each presentation, the participants propose different target systems, quality criteria and justifications for relevance. Following a joint debate, an initial model scope is agreed upon. It is important to note here that during multiple cycle iterations, alterations of the model scope can lead to modifications in the presentation for each participant. This can further propagate into changes at the level of personal knowledge and belief, changes in how direct observation is interpreted or changes in data requirements and analysis --- which can require a new data collection process. \\
        
Once the model scope has been constructed, the model will transition to the next phase of the cycle, namely the candidate model construction phase. This transition happens via the processes of the translation zone, which are explained in greater detail in Section~\ref{sec:ds_met}. Briefly, this ensure that a consistent conceptual representation of the heterogeneous system under study is being constructed, based on the distributed systems metaphor and taking into account all the elements of the model scope.  \\
        
\begin{itemize}
    \item[] \textbf{Phenomena}: the phenomena under study. 
    \item[] \textbf{External knowledge \& beliefs}: the knowledge and beliefs of the participants in the modelling task about or related to the phenomena under study. These can be based on personal experience, cultural and societal norms or different forms of education and are different from one person to another. 
    \item[] \textbf{Phenomena Presentation}: initial conceptualisation of the phenomena constructed after the start of the modelling task and based on direct observation and inference, prior knowledge, beliefs \& available data. Given the direct influence of the beliefs, this is again different for each participant. 
    \item[] \textbf{Data}: the available data assets regarding the phenomena. Can be both qualitative and quantitative and recorded before or during the modelling task. Although data assets are static --- they do not change after being recorded --- they are still influenced by the external knowledge and beliefs of the recorders or by the recording mechanism itself. 
    \item[] \textbf{Target System}: conceptual representation of an area of study in which parts of the phenomena are manifested. Multiple ones can exist, but they must be clearly stated. For example, a model of organisational recovery under ransomware could have the following target systems: fleet of devices, network, external storage, ransomware behaviour, user movement. 
    \item[] \textbf{Model Scope}: the list of target systems, their relevance to the studied phenomenon and the set of criteria that deem a representation of a target system as `good enough'. These can be related to both structure and behaviour: for example, how close should be the implemented representation of a network to a real network --- would routing and DNS resolution be implemented explicitly? --- or, how close should be a pattern of network congestion obtained from the model execution to one recorded in real world data.  In the spirit of co-design, all three are dynamic but should only be updated after an iteration through the translation zone or the complete cycle. 
\end{itemize}

\subsubsection{Candidate model construction}

The main goal of the Model Construction phase is to produce a functional heterogeneous model and to use it to obtain results via a form of execution. The forms of execution are different based on the nature of the component: conceptual and formal components can produce consequences directly by deduction or results via exemplification; executable components can only produce results, computationally by code execution or physically by performing operations on the physical model. It is worth emphasising here that the configuration of the model regarding the metric detailed in Section~\ref{subsec:model_qual} can change during cycle iterations. For example, conceptual model components can be implemented or formalised. Formal model components can be implemented or conceptualised. Executable model components can be formalised or conceptualised. The decision-making process used for changing the nature of the component is usually influenced by the modelling goals and the environment in which the model will be further deployed and used. \\
        
After the construction of the model, three phase transitions are possible: towards domain exploration via the second area of the translation zone, towards model use via deployment or towards model consequences derivation. Moving back towards domain exploration indicates that during the construction of the model, a need for an update in the model's scope has been identified. Moving towards model use carries the implication that all the representational criteria have been met and that the model is considered ready for use. Lastly, a move towards model consequences derivation requires different operations based on the nature of the model component: consequences can be drawn directly from for formal or conceptual components via deduction, or indirectly from any kind of component that has produced results via analysis, interpretation and then reasoning about such results.    \\
        
\begin{itemize}
    \item[] \textbf{Conceptual Model representation}: conceptual model including all target systems, structured by the distributed systems metaphor, agreed upon by the participants and clearly expressed. \\
    \item[] \textbf{Formal Model components}: the model components expressed using formal constructs such as systems of equations or logical formulae. \\
    \item[] \textbf{Executable Model components}: the implemented model components which manifest themselves in either a physical or computational environment. \\
    \item[] \textbf{Results}: the outcomes produced by a form of execution of the model. For executable components, the results of code execution or performing actions on the physical model. For formal or conceptual components, the traditional meaning of execution does not have a direct correspondence. However, formal and conceptual model components can be used to produce results directly --- not consequences in this case ---  via a process of exemplification. \\

\end{itemize}

\subsubsection{Model consequences derivation}

As the name suggests it, the purpose of this phase is to obtain a set of consequences about the model in a consistent format. For ease of understandability, this should be natural language, structured by the distributed systems metaphor that we detail in Section~\ref{sec:ds_met}. The consequences can be obtained via deduction from formal or conceptual components, or the analysis, interpretation and reasoning about the results or of the process of execution itself. To obtain the consistent format, the executable or formal consequences must be translated into natural language. Afterwards, a verification process is used to ensure no contradictions or incompatibilities can be found between the consequences. We note here that since up to this point, the consequences have not been reflected back into the domain, the verification process cannot ensure that all the issues have been determined. Only that there is no mismatch between the model components. Therefore, after multiple iterations through the verification process, the model might contain no mismatch between components but still not behave as expected. \\
        
Two cycle transitions are possible in this state: backwards toward candidate model construction or forward towards domain consequences translation. For the first one, if contradictions or incompatibilities are found during the verification process, the cycle backtracks to the previous phase and the model components are updated accordingly. If the verification process terminates successfully the model can transition towards the domain consequences translation phase by interpreting the consequences in the context provided by the domain. \\
        
\begin{itemize} 
    \item[] \textbf{Formal Model consequences}: the set of consequences obtained from the formal model components by using a form of deductive reasoning. These are usually expressed using a formal or semi-formal language. 
    \item[] \textbf{Executable Model consequences}: the set of consequences obtained from the analysis and interpretation of results or of the process of execution itself. 
    \item[] \textbf{Conceptual Consequences representation}: the set of all model consequences, in a clear natural language form. These can be obtained either by direct interpretation of the conceptual model components, or via translation from executable or formal consequences. It is of relevance here that the conceptual model consequences are not yet reflected back to the domain, but maintained at at the level of the model. 
    \item[] \textbf{Verification}: the process of determining whether or not contradictions or incompatibilities are present between the consequences. To ensure consistency, all the consequences must be expressed in the same language. For the procedure to be understandable to all the modelling participants without additional explanation, the consequences should be expressed using natural language. However, in the case of extremely large models, additional formalisation, implementation and execution procedures might be used for automatic formal verification. Needles to say, in such cases, the identified contradictions or incompatibilities should be translated back to natural language and further analysed. 
    
\end{itemize}

\subsubsection{Domain consequences translation}

The main aims of this phase are the interpretation of model consequences --- in our case the conceptual consequences representation in natural language --- to domain consequences and the validation of such consequences with respect to the domain, via the model scope. Model consequences are interpreted in the domain usually by contextualisation with the environment: for example, an executable consequence such as the model cars moving at an average speed of 30~mph is now interpreted as real cars moving on real roads at 30~mph. A formal consequence such as elements of type A not being included in set T of regular traffic participants can now be understood as ambulances not having to stop at traffic lights --- this example is trivialised, since we do not fully explore all the implications of being part of set T. After this contextualisation step, validation procedures are used to determine whether or not the quality criteria from the model scope have been achieved. \\
        
A single cycle transition is possible in this state: towards the domain exploration phase. However, the result of the validation process determines if subsequent cycle iterations are required: if mismatches between the translated consequences and the model scope are identified, updates will be required in the next iteration through the translation zone, which will then be further propagated at the level of the constructed model components; if not, the next iteration through the translation zone can be used for optimisation purposes or simply for the participants to decide if the model is considered ready to be used. 
        
\begin{itemize}
    \item[] \textbf{Domain consequences}: formal model consequences, executable model consequences and conceptual consequences derived directly from the conceptual model representation translated at the level of the domain. \\
    \item[] \textbf{Validation}: the process of comparing the translated domain consequences with the model scope and ensuring that the fidelity criteria for the representation have been met. \\
\end{itemize}

\FloatBarrier 
\newpage






\section{Trading zones, distributed systems \& translation zones}\label{sec:ds_met}

As discussed in previous sections, the act of modelling is a collective action that involves different actors that partake in it for different reasons, possess different sets of skills, knowledge and beliefs about the object of modelling and are able to construct different representations of it. The above described co-design cycle represents an attempt at structuring this process at a high level of abstraction. However, the reader might notice after a careful analysis that the cycle on its own does not detail on how to manage and align the plethora of different interpretations that the participants create. In this section, we will address this by the following approach: firstly, we describe what a trading zone is and some of the benefits that can be obtained by using this conceptualisation at the level of the design stages in the modelling cycle; secondly, we introduce the idea of a distributed systems metaphor as structuring element to guide the above stages toward a specific type of trading zone; thirdly, we showcase the resulting translation zone and explain how it integrates with the co-design cycle.
\subsection{The trading zone}\label{subsec:trading_zone}

As coined by Peter Galison in~\citep{galison1987, galison1990, galison1997i}, the term trading zone describe situations where people from different disciplines or cultural backgrounds collaborate and communicate, despite having distinct languages, methods, and practices to achieve a high degree of understanding across a multidisciplinary domain of inquiry. The idea has been inspired by anthropological practice and was used by Galison to describe how physicists focused on different paradigms managed to jointly collaborate with engineers to construct the radar and particle detectors. However, its degree of generality is higher than that. In the author's own words: `Two groups can agree on rules of exchange even if they ascribe utterly different significance to the objects being exchanged; they may even disagree on the meaning of the exchange process itself. Nonetheless, the trading partners can hammer out a local coordination, despite vast global differences. In an even more sophisticated way, cultures in interaction frequently establish contact languages, systems of discourse that can vary from the most function-specific jargons, through semispecific pidgins, to full-fledged creoles rich enough to support activities as complex as poetry and metalinguistic reflection'~\citep{galison1997i}. 

Building on Galison's metaphor,~\citep{collins2007} further extend the notion and classify different types of trading zones by analysing the nature of the cooperation between participants and of the resulting culture. This results in the conceptualisation of four new types of trading zones: inter-language, subversive, enforced and fractionated. For the purpose of our paper, we will focus our attention on the fractionated trading zone. A more in depth analysis of each of the zone types and relationships with constructed and deployed models is deferred to further work.

The primary descriptors of a fractionated trading zone are high collaboration, heterogeneity of resulting culture and fractions of culture as medium of interchange. Given the materiality aspect of the fractions of culture, the resulting trading zone can be based either on boundary objects or interactional expertise. 

As described in~\citep{star1989}, boundary objects are an `analytic concept of those scientific objects which both inhabit several intersecting social worlds and satisfy the informational requirements of each of them. Boundary objects are [...] both plastic enough to adapt to local needs and the constraints of the several parties employing them, yet robust enough to maintain a common identity across sites'. Furthermore, they can be both abstract and concrete as long as they remain recognizable at the level of each intersecting social worlds.

Interactional expertise, on the other hand, represents a form of expertise in understanding and using the language, concepts, and practices of a particular domain or community without possessing practical or hands-on skills in that domain. In~\citep{collins2002third}, this is expressed as `enough expertise to interact interestingly with participants and carry out a sociological
analysis', but the sociological analysis element should be interpreted as specific to their topic of inquiry --- to maintain generality, we interpret it as an analysis of the topic. Interestingly~\citep{collins2007} also describes interactional expertise as a linguistic complement of boundary objects developing through linguistic socialisation. 

The above described concepts can be easily observed at the level of modelling. In this setting, perhaps the most relevant aspect to notice is that the model to be constructed acts as boundary object --- the nature of heterogeneous models does not contradict~\citep{star1989}, as long as a common identity for the model is maintained by the participants. Since the notion of heterogeneous model inherently implies a multidisciplinary domain, we can argue that the process of designing and constructing such models can be viewed as a fractionated, boundary object trading zone. 

However, as~\citep{collins2007} notes, trading zones do not remain stable, but tend to evolve and transition to different types over time. In our case, a transition towards a subversive or enforced trading zone would not be desirable: the former implies one culture overwhelming the other participants' which brings the risk of biases being more easily introduced in the model, whereas the latter usually implies a lack of cultural interchange and therefore a reduction in multidisciplinary understanding of the phenomena under study. Furthermore, a transition towards an inter-language trading zone, although desirable in later stages of the modelling process due to the increased collaboration, is most of the times impossible due to time constraints and differences in goals of the participants. Stakeholders, users and domain experts usually partake in modelling activities alongside other projects and do not inherently aspire to become modellers. Last but not least, attempting to maintain a classic boundary object trading zone also has its disadvantages, since the participants would essentially work `in silo' and not share their knowledge and expertise --- this can be very problematic when attempting translation and implementation of model components and lead to biases remaining hidden and conserved. 

Therefore, the following trading zone evolution trajectory could be considered desirable: starting as a fractionated, boundary object trading zone --- with or without initial coercion in the form of `encouragement' during the start of the project --- increasing the degree of collaboration and interest of the participants regarding the topic up to the point of development of interactional expertise to also facilitate knowledge sharing and then striving towards an inter-language trading zone by attempting to increase the homogenity of the resulting culture via new cultural tools as described by~\citep{collins2007}. One such conceptual tool is represented by the distributed systems metaphor that we present in the following section.



\subsection{The distributed systems metaphor}\label{subsec:ds_met}

The expansion of interconnected network systems gave rise to the formulation and advancement of distributed systems theory in the field of computer science. Under this paradigm, a distributed system can be thought of as a collection of physically independent entities that communicate and coordinate with each other to achieve a common goal. The entities have been traditionally related to computing, so technological in nature --- in the case of heterogeneous systems, this is no longer the case ---, the communication is performed across a network, the coordination is realised via scheduling algorithms and the common goal is represented by the service the system is supposed to provide to its users. 

While this theory has been historically focused on computer systems, its principles and core components can be applied more broadly as a useful metaphor for understanding various types of systems, including ecosystems. In previous work such as~\citep{caulfield2021eng}, we describe the ontology of the distributed systems metaphor as containing the following primary concepts: 

\begin{itemize}
    \item[] \textbf{Location}: Distributed systems inherently involve the idea of multiple locations interlinked with each other. These locations may represent physical entities --- such as rooms connected by passageways ---, logical entities --- like addresses in computer memory connected by memory pointers ---, or abstract entities --- for instance, the conscious and unconscious areas of cognition interlinked by dreams. Formally, locations are represented by directed graphs. \\

    \item[] \textbf{Resource}: Resources exist at locations and can be moved between them according to the locations' connections. Generally, they can be used to represent anything that can be manipulated by a process --- data, physical and abstract objects, people, etc. For example, computers in rooms, software programs loaded in computer memory or memories in the unconscious can all be viewed as resources. Formally, resources can be represented as elements of a pre-ordered, partially commutative monoid, as provided by the semantics of bunched logic~\citep{o1999logic,CMP2012}. \\
    
    \item[] \textbf{Process}: Processes are used to represent collections of actions that manipulate resources in an either sequential or parallel manner. Such manipulations can include moving resources from one location to another, generating or removing resources from locations, composing resources, altering the internal structure of a resource via decomposition, and so on. Formally they can be represented --- in the spirit of Milner's SCCS~\citep{milner1983} --- using a monoid of basic actions, a grammar of process terms to describe process interaction and a partial modification function illustrating the co-evolution of processes and resources under actions.   \\
    
    \item[] \textbf{Environment}: The systems we model are not isolated entities; instead, they interact with environments that we choose not to represent in detail. Such environments are used to depict the worlds outside of the system of interest and the interactions between such worlds and the system. Semi-formally, environments can be represented stochastically by capturing the incidence of events they generate towards the model. For example, given an organisational physical security model, the arrival rate of agents at the entrance that marks the outer boundary of the model may be captured using a negative exponential distribution~\citep{caulfield2016mod}.   \\
    
\end{itemize}

Because of scope and space constraints, we shall not dwell on the formal theory developed around this concepts in the areas of computer science, logic or mathematics any further. For more precise descriptions, the reader can consult~\citep{o1999logic, CMP2012, caulfield2016mod, caulfield2021eng}. In the following subsection, we shall focus on showcasing some of the properties provided by this conceptualisation in a modelling context. \\


\subsection{Properties of The Distributed Systems Metaphor}\label{subsec:ds_met_prop} 

The utility of a metaphor lies not solely in its literal accuracy, but rather in the extent to which it enhances understanding, facilitates communication, and elicits deeper insights. While the direct correspondence between the metaphor and its subject may be imperfect, the benefits derived from its usage outweigh mere literal precision. Below, we outline the benefits of using the distributed systems metaphor approach in the context of modelling heterogeneous systems:

\begin{itemize}
    \item[] \textbf{Generality}: This conceptualization offers a set of ideas applicable for modeling virtually any type of system. Historically, notions of systems can be traced back to Plato, Descartes or the development of cybernetics in the 19th century --- which specifically included various types of systems such as ecological, technological, biological, cognitive and social systems. For a detailed account, see~\citep{franccois1999s}.
    
    \item[] \textbf{Recognizability}: Based on the generality aspect, there is no surprise that the above concepts have been studied and can be mapped across a wide array of scientific areas. For example, definitions of a notion of process can be easily found in the fields of business and management, economics, law, psychology, philosophy, physics, chemistry, medicine, computer science, mathematics, logic and others. It is worth noting that although these definitions are specialized for their corresponding areas --- and expertise from many such areas might be required to model a heterogeneous system --- they still maintain the core idea of a collection of actions that leads to a form of result.
    
    \item[] \textbf{Scale-freeness}: The above described set of concepts can be used for constructing representations at any level of abstraction. For example, a location could be an area in computer memory, a room, an office building, an entire city. Similarly, a resource could be a network data packet, a configuration file, an unit of energy, an amount of money or an entire fleet of IoT devices.  
    
    \item[] \textbf{Formal Properties}: The formal theory underlying these concepts describes three extremely useful properties for a modelling formalism: \emph{composition}, \emph{substitution} and \emph{local reasoning}. \emph{Composition} describes a structured way of constructing model components by connecting or combining smaller sub-components that inherently helps with managing system complexity --- for example, the organisational ransomware recovery model described in Section~\ref{sec:case_studies} is composed of four underlying sub-models: a storage server model, a network model, a physical organisational model and a ransomware behaviour model. \emph{Substitution} illustrates the necessary and sufficient structural conditions for a model component to be replaced with another. This ensures that a replacement of model sub-components would not bring the overall model into a misconfigured state and can be particularly relevant when adjusting the level of detail, exploring different designs for parts of the model, or replacing a portion of the environment with a specific model. For example, in the case of the emergency trauma unit model in Section~\ref{sec:case_studies}, the environment generating patients based on arrival rates could be substituted with an explicit triage model. Last but not least, \emph{local reasoning} refers to the process of making inferences or drawing conclusions about a specific part of a model without considering the entire model. The main benefit of this lies in the ability to analyze the properties of a particular component in a model's decomposition without the necessity to consider other components, except in relation to their interfaces or connection points with that specific component. A precise, formal description of these properties can be found in~\citep{caulfield2021eng}. 

    \item[] \textbf{Implemented Tools}: The foundational principles underlying the distributed systems metaphor have been utilized to create practical frameworks for model construction. One of the earliest attempts at implementation was Simula~\citep{dahl1966simula}, an Algol simulation framework that primarily emphasized processes. Subsequent implementations like Birtwistle's Demos~\citep{Bir79,Bir81} or Gnosis~\citep{CMP2012} aimed to expand the conceptual toolkit to include elements like resources or locations. Furthermore, the authors propose a newer implementation in the Julia Language --- specifically the publicly available SysModels package as described in~\citep{juliacode} --- incorporating more of the above described properties. 

    \item[] \textbf{Identity Conservation}: Constructing models using this approach satisfies the main criterion for the functioning of a trading zone, as shown in~\citep{collins2007}. As long as the modelling participants have a common understanding of the concepts --- process, resource, location, etc. ---, a common identity for model components can be maintained in relation with them. We note here that maintaining this common sub-component identity implies first constructing it collaboratively. This facilitates the development of interactional expertise and knowledge sharing about the underlying sub-components: when debating the possible configuration of model sub-components, the participants inherently construct a personal understanding of them which can be directly shared because the set of concepts acts as an in-between language. 
    
\end{itemize}

\subsection{Comparing the two metaphors}\label{subsec:comp_met}
In previous subsections, we have described the distributed systems metaphor and made arguments for its usefulness in a modelling context. We have also determined that it would be desirable for the modelling processes described by the co-design cycle to follow a specific trading zone trajectory: from fractionated towards inter-language. This implies homogenising the resulting culture via new cultural tools~\citep{collins2007} and is associated with the development of in-between vocabularies. We argue that the distributed systems conceptualisation, can be considered as such a cultural tool, and could act as an in-between vocabulary, but employing it in a trading zone context should only be done if the two metaphors are essentially describing similar situations. 

The specific choice of a distributed systems conceptualisation is not arbitrary: a certain similarity exists between trading zones and distributed systems. Based on the definition of trading zone from Section~\ref{subsec:trading_zone}, we illustrate the similarities with distributed systems metaphor below:

\begin{itemize}
    \item[] \textbf{Entities}: The primary constitutive elements of a trading zone are the people partaking in the `trading'. For distributed systems, the subsystems are composed of a mixture of people, technology related components and policies. In both cases, the people can have different cultural backgrounds and areas of expertise; the technology components can be developed based on different scientific traditions; the policies might affect any of the components. 
    \item[] \textbf{Interaction}: The trading zone metaphor, being specifically focused on people, describes the main attributes of the interaction between parties as communication and collaboration. Similarly, the sub-components of a distributed system must communicate and coordinate their operational activity. The situation is more complex in the case of distributed systems, because the underlying entities can be different in their nature: people might have to interact not only with other people, but also with policies and technology based components.
    \item[] \textbf{Language, methods \& practices}: Based on the cultural and expertise differences, the participants in a trading zone might speak different languages and adhere to different sets of methods and practices. The same can be said for the distributed systems case, with the note that this is manifested also for technological components and policies. For examples computers might run different operating systems, execute code in various programming languages, vary in development and testing methodologies and have very specific sets of policies. 
    \item[] \textbf{Goals}: In both cases, multiple sets of goals exist. For example, trading zones exist because the participants obtain something by partaking in them --- knowledge, economic benefits, reputation, etc. --- but at the same time have an overall goal --- for example to produce a boundary object --- which would be hard to achieve in a different setting. At the level of a distributed system the same can be observed: people have different incentives for being part of the system --- most of them related to being part of the organisation that owns the system --- technology components are usually designed for specific purposes, but their practical use can differ from that purpose and, policies can sometimes even serve purposes that are not directly related to the system, but to some higher level goal --- for example, a global organisational spending policy might drastically reduce design options for a system, without considering the implications for its construction. Yet, all the different components serve a common goal, to produce a service or product which is significantly harder or even impossible to produce in isolation.
\end{itemize}

Based on the above descriptions, we can argue that trading zones and distributed systems manifest a high degree of similarity at a conceptual level. Therefore, using a distributed systems conceptualisation in a trading zone context is not far fetched. 

\subsection{The translation zone}\label{subsec:translation_zone}

In the previous subsections, we have illustrated the compatibility between the two metaphors and some of the benefits of employing the distributed systems conceptualization in a modelling context. Further, we shall explicitly describe how these ideas can be integrated with the co-design cycle from Section~\ref{sec:co_design}, at the level of a translation zone. 

First of all, the translation zone represents the area of the co-design cycle that links the domain exploration and model construction phases. More precisely, it can be viewed as a bi-directional set of processes, whose purpose is to align the different goals, epistemic beliefs, practical interests and expertise of the modelling participants, with the help of the distributed systems metaphor, in an attempt to produce a co-evolution of model scope --- target systems, representation criteria, relevance --- and constructed model and, facilitate knowledge sharing and the development of interactional expertise, by design.

As show in Figure~\ref{fig:translation_zone}, the processes of the Translation Zone can be separated into three different areas:

\begin{figure}[hp]
    \centering
    \includegraphics[scale=0.63]{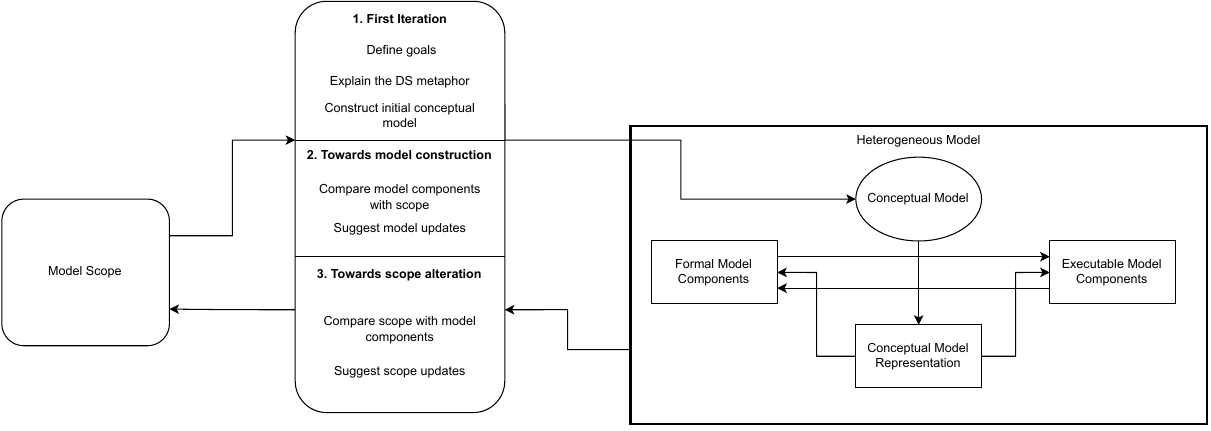}
    \caption{The translation zone}
    \captionsetup{justification=centering}
    \label{fig:translation_zone}
\end{figure}

\subsubsection{First iteration}

The first iteration through the Translation Zone, from the Domain Exploration towards the Model Construction phase can be seen as an initialization procedure. Starting from a relatively generic initial model scope, the participants aim to construct an initial conceptual representation of the heterogeneous model that includes all the relevant target systems. Abstractly, this can be seen as a dual process of reasoning and translation: given the nature of the available information about the systems, the reasoning process can be deductive, inductive or abductive followed by a translation towards the distributed systems conceptualization. Furthermore, we detail the specific steps to be undertaken:

    \begin{itemize}
        \item[] \textbf{Define goals}: The overall model goal and the goals of each participant should be clearly defined, noted and agreed upon. 
        
        \item[] \textbf{Explain the metaphor}: The responsibility of ensuring that the participants understand and are familiarised with the concepts employed in the distributed systems metaphor lies with the modeller. To facilitate this, the modeller should consider the background, knowledge, and expertise of each participant and tailor the language used and level of detail accordingly. Furthermore, the use of analogies, real-world examples and a positive, open atmosphere where questions and feedback are encouraged can greatly improve the process. 

        \item[] \textbf{Construct initial conceptual model}: For each target system included in the model scope, a conceptual representation must be constructed. For that to be possible, the participants must first identify what elements will be directly expressed and at what level of detail and, note down justifications for the omitted ones and for any underlying assumptions. Then, they must decide how to map each element to the distributed system concepts --- what is a resource, what is a process, what is a part of environment, etc. We note here that multiple possible mappings can exist: for example, a building could be considered both a resource and a location, or an employee could be seen as a resource or an implied entity that starts a series of different processes --- nevertheless, the decision is highly situational. Lastly, the participants suggest a direction for the construction of each model component --- conceptual, formal or executable --- by explicitly taking into account the model goals and the future deployment environment. 
    \end{itemize}

\subsubsection{Towards model construction}

Similarly to the initial iteration, the goal of this phase is to produce a conceptual model representation according to the criteria in the model scope. However, the main difference is that model components have already been constructed in some form so, the stage can be understood overall as aligning the components with the scope. In practice, the following steps should be considered: 

     \begin{itemize}
        \item[] \textbf{Compare model components with scope}: To ensure that the model is being developed in the collectively agreed upon direction, each model component must be compared with its associated scope element. That means checking that both the conceptual representation of the component and its practical implementation --- if it exists --- are adhering to the representation quality criteria. For example, if the target system is a network, the quality criteria is based on recorded real-world data about the target and an architectural diagram and, a previous decision has been made to construct this component as an executable simulation, then the participants must ensure that the results of the simulation illustrate a network behaviour closely resembling the real-world target --- the degree of closeness should be specified --- and that the distributed systems conceptualisation of the network matches the architecture of the real network.   
        
        \item[] \textbf{Suggest model updates}: Based on the above comparisons, areas of improvement can be identified and corrective actions are proposed, in line with the scope. For example, the simulated network's behaviour might not closely resemble the real network, but the verification processes might not identify any issues. In such a case, additional domain understanding should be required: interaction with engineers and network administrators could perhaps reveal that the network performance is influenced by additional constraints --- perhaps backups and maintenance are conducted during certain hours, therefore reducing the available bandwidth, but this was not detailed in the original recorded dataset. In light of the discovery, the participants must now decide whether to explicitly structure the backup and maintenance processes and then implement them, or perhaps implicitly reproduce the behaviour by altering network control parameters.  

    \end{itemize}

\subsubsection{Towards scope alteration}

As the name suggest it, processes in this phase deal with the other aspect of the co-evolution, namely ensuring that the quality criteria and model scope are still relevant to the model. 

    \begin{itemize}
        \item[] \textbf{Compare scope with model components}: Similarly to the first step above, the participants must first analyse and compare the model components and scope. However, this time, their purpose is focused on identifying possible areas of improvement in the scope. For example, in the above case, the network target system can be seen as an area of improvement at the scope level.  
        
        \item[] \textbf{Suggest scope updates}:  As expected, the identification of areas of improvement is followed by actions that alter the scope in that direction. For example, if the participants decide to explicitly implement the additional network processes, additional structural quality criteria must be added to the model scope. However, this is not the only case when the scope could require updates: perhaps due to performance constraints, a model component shall be represented at a higher level of detail, to reduce the number of instantiated entities; opposingly, another component might require a translation from formal to explicit implementation, in an attempt to increase the model understandability. Needless to say, any such changes should be collectively agreed upon, documented and reflected in the updated scope.  

    \end{itemize}

\newpage
\section{Case studies}\label{sec:case_studies} 

In the interest of exemplification, we briefly introduce three different security related models constructed using the distributed systems metaphor and co-design approach, in previous work: a physical data loss model, an organisational recovery model under ransomware attacks and a trauma room surge capacity model. We describe their goals, internal structure and representation choices and illustrate how each of the models focuses on different aspects of the metaphor: the data loss model heavily focuses on physical locations, the ransomware recovery one on processes and the trauma unit one on resources. Subsequently, we explain the iteration of the co-design cycle steps within these three specific contexts and discuss the challenges encountered. Due to space constraints and scope, we do not include elements such as precise parameters and testing suites, analysis of the produced results or the actual code used. 

\subsection{Physical data-loss model}




The first model we introduce here was developed in~\citep{caulfield2016mod} and aims at assessing the impact of different physical security policies with respect to data loss at the level of a small-sized organisation. This is a simple model that illustrates the use of the distributed system metaphor approach and the role of the co-design cycle in a relatively uncomplicated setting.

We must note that this is not the first model ever constructed using a version of the distributed systems metaphor. For example, the executable models of~\citep{Bir79, Bir81} are based on a system representation including processes and resources, but without an explicit characterization of the nature of systems such as that 
provided by the distributed systems metaphor, and without explicit 
conceptualizations of locations, environments, or interfaces. 
Furthermore, the authors have continued to develop the conceptualisation in a series of works such as~\citep{CP09,CMP10,CMP2012,AP16}, and consider applications in, for example~\citep{beautement2009m, cpw14, PymShiu:IISP, baldwin2021m}. A previous implementation of these ideas, Gnosis~\citep{CMP2012}, has been used in significant commercial applications~\citep{BaldwinEtAl2012,BeresEtAl2008,BPS2010} derived from an industry-based research project \citep{security-analytics}.

\subsubsection{Structure \& representation choices}

\begin{figure}[hbt!]
    \centering
    \includegraphics[scale=0.5]{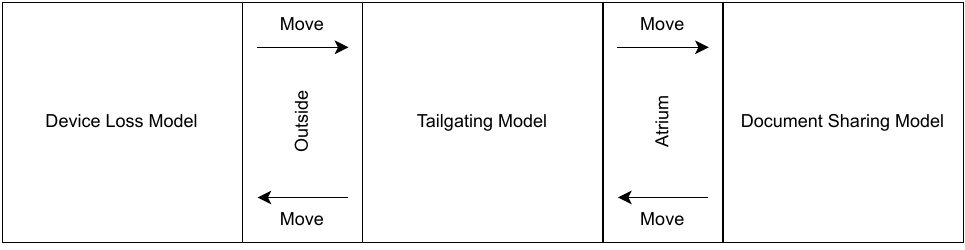}
    \caption{The simplified physical data loss model}
    \captionsetup{justification=centering}
    \label{fig:sp_datal_mod}
\end{figure}

As previously stated, this model attempts to provide a better understanding of the possible impact of physical security policies in relation with data loss. In order to capture the main elements related to these phenomena, we focus on representing the locations where physical data loss can occur: in the office, if external access is possible, or in transit. Structurally, this leads to three different sub-models, depicted in Figure~\ref{fig:sp_datal_mod}.

\FloatBarrier 

The device loss model, situated on the left side of the diagram, acts as an abstract representation of the areas outside an office building where data assets can be lost by employees in transit. Its operation can be described succinctly: while commuting to or from work, employees face the risk of losing the devices they carry, potentially containing confidential data. The quantity of confidential documents stored on these devices is influenced by the behavior of employees in the document-sharing model. 


The tailgating model is used to illustrate the physical boundary of the organisation and assess the ability of an external attacker to traverse it to gain access to internal data assets. In practice, employees are required to present an ID card resource to gain access to the office via a security door location where the ID verification process takes place. When employees forget their cards, they are faced with a decision: either wait in line at the reception desk to obtain a temporary ID card for the day, or attempt to tailgate through the door. Employees who have successfully passed through the security door then observe if others attempt to tailgate behind them. In this scenario, the employee has two options: either ignore the tailgater and proceed directly to the office, or confront the tailgater and redirect them back to the reception area. 



The document-sharing model examines how employees behave within the office environment when faced with decisions about sharing confidential documents among themselves. Typically, employees share documents through a shared drive that limits access to authorized personnel. However, this system often experiences downtime, requiring employees to resort to alternative methods for document sharing. Within the model, three options are available. Firstly, employees can utilize a global share accessible to all personnel. Secondly, they can opt to email the documents directly to recipients. Lastly, employees may choose to use portable media like CDs or USB sticks to share data. Each option presents its own drawback: documents on the global share are accessible to all employees, emailed documents end up on devices carried to and from work, and portable media left lying around the office poses a risk. Furthermore, attackers roam the office, collecting any abandoned portable media resource they encounter.

Given the above setting, we can now present the explicit mapping between the desired entities and phenomena to be modelled and the concepts of the distributed systems metaphor from Section~\ref{subsec:ds_met}, at the level of the composed model.

In terms of \emph{locations}, the model is relatively simple: most locations are physical and are used to represent areas such as the employees' home, public transport, private transport such as cars, an area outside the main building's lobby, and internal areas such as the lobby, entryway, atrium, or actual office. 

At the level of \emph{resources}, the situation is similar: the concept of resource is being used to represent physical items where confidential information might be stored, including devices such as mobile phones or laptops, cds, USBs, or paper documents. Furthermore, the ID cards required by employees to enter the building are also resources.

With respect to \emph{processes} the model can be viewed as slightly more complex. Processes are being used to represent a wide array of activities including travelling, queuing for a temporary ID badge, working, observing, challenging, ignoring, or performing tailgating, deciding how to send confidential documents, and then sending them, or searching for, or loosing data assets. Additionally, each process is associated with an agent symbolizing a regular employee, a security guard, or an attacker. Agents are constructed as a bundle of a resource, a process and multiple locations: the resource signifies the physical position of the agent within the model. The process is used for relocating this resource within various model locations, while also engaging with other resources as required. The positions associated with the agent are utilized to represent concepts such as possession or memory. For instance, to simulate an agent acquiring another resource, the agent's operation would relocate that resource to the position representing items being carried by the agent; conversely, releasing the resource would return it to a physical position within the model. 

Last but not least, \emph{environments} are used to represent areas of the model that are not conceptualised or constructed in detail. For instance, given the model goal, the explicit activities performed by an employee at home outside office hours or by an attacker outside an attack timeline are not relevant. Because of that, environments initialised with probability distributions are being used to start the processes of employees travelling to work or for attackers arriving in the main building.

\subsubsection{Co-design}

Having briefly explored the structure of each underlying sub-model and the associated representation choices, we now turn to illustrating the co-design steps involved. However, we first again emphasize that this initial example is quite simple and should be viewed as an educational illustration for employing the distributed systems metaphor, rather than a detailed explanation of the use of explicit co-design. It is noteworthy to mention that the development of a co-design theory for heterogeneous modelling, at least in the sense shown in Section~\ref{subsec:co_design_cycle}, was preceded by the construction of the physical data-loss model. 

Nevertheless, the setting of the model still provides us with the opportunity to illustrate how the co-design cycle might have been used in this case, as a thought experiment. We limit the description to the domain exploration phase of the co-design cycle due to space constraints and reduced scope. The other two models to be presented in the following subsections will be used to detail aspects of the co-design process not present in the physical data-loss model.

In this context, the domain exploration phase can be seen as heavily influenced by the composition of the modelling team which was composed of two modellers with a great amount of expertise in information security, but no other stakeholder or user. 

The phenomena to be studied are all relevant to the concept of physical data loss: physical data loss can only occur if an attacker can physically obtain access to a data asset. Assets can only be found in the office or in transit if lost --- we do not consider scenarios involving theft from residential areas --- and if they are in the office, the attacker must also be there. Therefore, it is reasonable to assume that the phenomena to be studied include transiting from home to the office, possibly losing data assets on the way, gaining access to the building, performing work, deciding how to share assets, deciding whether or not to tailgate into the building or to confront an attacker and so on. 

Similarly, the entities involved in these phenomena are the agents --- regular employee, security guard, attacker --- the resources themselves --- data assets, ID badges --- and the locations where may be found --- home, transit, office, lobby, etc.. 

With respect to data, the only source used was probabilistic: a negative exponential distribution was used to model the arrival rate of employees at the entrance to the office building. Other aspects related to the organisational security posture or attacker behaviour such as the probability of an employee or security guard to challenge an attacker, the probability of losing a device in transit, the attackers' arrival rate, or probability to discover a data asset were controlled through model parameters and used to generate different scenarios. 

In light of this, we can view the model scope as containing 3 main target systems, closely matching the sub-models in Figure~\ref{fig:sp_datal_mod}: a transit system, an office entrance system and an office internal space system. Their relevance is clear: they represent the areas in which physical data loss can occur. Given the lack of physical deployment and the exploratory rather than practical nature of the model, the representation criteria for the target systems are very simple: the employee arrival rates, device loss rates, successful tailgating, successfully discovery of assets, challenging an attacker and so on must resemble possible real-world occurrences reasonably close. In a practical, organisational case, this would have been very different: quantitative historical data about arrival rates, previous incurred attacks, office document disposal and storage policies, and past stakeholder experience can be used for a more detailed representation involving stricter quality criteria. 

The establishment of the model scope marks the completion of the domain exploration phase, at least for the current iteration of the cycle. Given specific goal of experimentation with a simulation model, there is no surprise that the translation zone is minimal --- conceptualizing the target systems using the distributed system metaphor did not require extended multi-disciplinary debates --- and the resulting model is heavily simulational. 


\subsection{Ransomware recovery model}

    
The subsequent model we aim to introduce here originates from yet unpublished, but currently under review work available publicly at~\citep{HPart2023}. The primary goal of the ransomware recovery model is to illustrate the impact of deploying different recovery methods at the level of a small to medium sized organisation affected by ransomware attacks of varying severity. This is a significantly more complex model than the physical data-loss one in terms of both structure and co-design: the represented entities are more varied, and multiple stakeholders have been involved in the process. Because of that, we will use this model for exemplifying areas of the co-design cycle such as model consequence derivation, domain consequences translation, and a translation zone iteration that altered the candidate model. 

\subsubsection{Structure \& representation choices}

\begin{figure}[hbt!]
    \centering
    \includegraphics[scale=0.5]{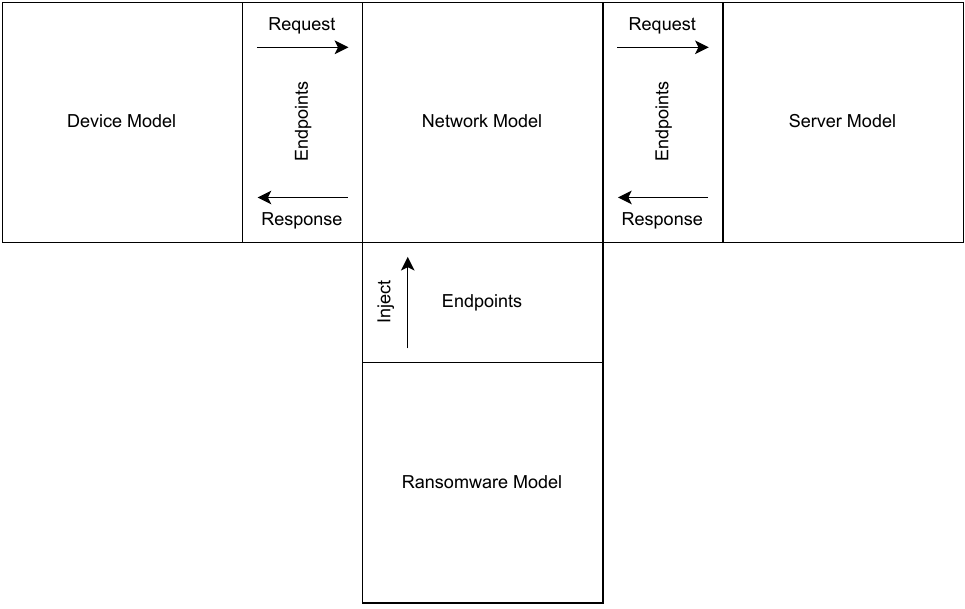}
    \caption{The simplified recovery model}
    \captionsetup{justification=centering}
    \label{fig:sp_rec_mod}
\end{figure}

Structurally, the model is composed out of four different sub-models, each depicting a relevant area for organisational recovery: a device model, a network model, a server model and a ransomware model. A simplified version of the composed model is shown in Figure~\ref{fig:sp_rec_mod}.

The device model stands out as the most intricate among the four, and is used to represent the physical structure of a medium-sized organisation. This is comprised of a series of physical locations ---  offices of varying sizes, home, hotels, coffee shops --- where employees can work on their devices if a network connection is available. Naturally, employees can move between locations, but the specific location they are present in influences their available resources and restricts their possible recovery choices: locations have different network bandwidth allocation and non-office locations do not have access to a physical help desk. An iteration through the model should be viewed as a temporal sequence of movement between locations, working, getting infected by ransomware and then performing recovery actions --- USB recovery, network recovery or embedded recovery or combinations of them.

The network model acts as the central representation for the organization's communication network and facilitates the interaction between all the models via the transfer of network packets. Functionally, its goal is to ensure network packets arrive at their correct destinations after an amount of time influenced by the network congestion. Structurally, it contains abstract locations symbolizing network endpoints that devices can access to connect to the network. These locations are interlinked, representing the network segments that actual network packets would traverse. Briefly, the model operates as follows: network packets arrive at endpoints, are transferred to an abstract transit location, and after an appropriate delay aligned with data size, network segment speed and congestion, dispatched to their respective destination endpoints.

\FloatBarrier 


The server model essentially depicts the storage area for recovery images which are requested by devices over the network, depending on the recovery type chosen in the device model. Its behaviour closely resembles the real interaction between a user and a server: it is used to determine if network requests contain valid recovery image requests and if so, to send the required images to their corresponding devices via the network.

The ransomware model is used to encapsulate the main aspects of the ransomware behaviour: the targets, the spread pattern and the infection timings. In practice, it contains processes for determining targets for infection and uses different probability distributions representing infection severity timings to inject ransomware packages on the network. After the injection procedure, the packets are handled by the network model, reach their targets, infect them and then the device model triggers the recovery processes. 

In light of the context provided, we can now display the direct correspondence between the entities and phenomena to be represented and the ontological components of distributed systems metaphor.

Regarding \emph{locations}, the model manifests relatively high diversity. For instance, some locations such as the network transit location are abstract, representing the area of the network where the time delays are being calculated before network packets are routed to their targets. Others, such as the network endpoints or the server storage location are physical, but contain data related resource. Finally, there are also physical locations like homes, offices, or other travel locations which are directly related with the employees and their devices. Each of the latter locations has a series of associated former ones: for example, a large office location has a wired and a wireless network connection endpoints, but a hotel only has a wireless one. 

Due to the relationship between \emph{resources} and \emph{locations}, a similar level of diversity manifests in the latter. For example, some resources are heavily data oriented: operating system images, image data requests, image responses, usb images, the recovery agent, or simply network data packets. Similarly, other resources like desktops, laptops or USBs are being used for storing or processing the above data resources. Lastly, help-desk admins are also represented as resources which may be necessary in the case of recovery for users without great technical expertise.

In terms of \emph{processes}, the situation is even more complicated. For example, a set of processes including choosing targets, determine attack timings, and infecting the devices via the network model are being used for representing the main characteristics of ransomware behaviour. Another set of processes is used at the level of the network model for processing data packets, transferring data between endpoints or to the transit location, or calculating timings for the network transfer. Yet, another set of processes is related to the devices: they can be used for performing different recovery options --- USB, embedded, network-based or mixed ---, request different help-desk assistance procedures, installation of recovery images or the transport of devices between locations. Lastly, a separate process is being used for calculating the timing delays introduced by the help-desk procedures required for completing recovery. 

With respect to \emph{environments}, the only use at the level of this model can be found in the Device Model, where an environment is being used for starting the process of device movement. However, in a simpler version of this model~\citep{baldwin2021m}, the entire Ransomware Model was replaced by an environment which started attacks based on a probabilistic, uniform distribution of attack timings.

\subsubsection{Co-design}


    



The construction of the organisational ransomware recovery model was the result of a collaborative research project between the authors of this paper and HP Security Lab Bristol which remains as of now yet unpublished, but available in~\citep{HPart2023}. As stated previously, this model is significantly more complex than the data-loss one, and provides use with the opportunity of exploring other areas of the co-design cycle. Specifically, we will focus on a translation zone iteration that altered the candidate model, and a derived model consequence, its interpretation in the domain, and the validation procedure which led to the discovery of an undesirable, possible real-world situation.  

The motivation behind it was twofold. Firstly, to better understand how different recovery techniques can be allocated at an organisational level, in an attempt to reduce the impact of ransomware attacks of varying severity. Secondly, to explore the suitability of the distributed metaphor and co-design approach in a technical, real-world security organisational setting. 

The modelling team included the following participants: a senior security research manager with a particular interest in ransomware who acted as main stakeholder and had extensive knowledge in both the security research and organisational areas, three modellers, each focused on formal, conceptual and simulation models, a security architect who was involved in the design and construction of the recovery technologies represented in the model and a help desk specialist. The interaction between members was realised through a series of meetings with different configurations. For instance, the security research manager and the three modelers met regularly --- on average, twice every two weeks --- and were involved in all aspects of the co-design cycle, including explicit analysis of the software code. We note here that during these meetings, the security research manager learned and started actively using the distributed systems conceptualization. However, the security research manager also conducted separate meetings with the security architect, and help-desk specialist in the interest of data collection, and better understanding of the recovery technologies --- some of which were constructed by the architect's development team --- help-desk employee behaviour and timings. The direct participation of the security architect and help-desk specialist in meetings involving all the participants was not possible due to time constraints and other organisational commitments. However, in later stages of the model development, one of the modellers was deployed in the organisational setting for six months, and interacted with them directly --- with the research manager, on a daily basis, and with the others, on a case-by-case basis --- for the purpose of validation and possible identification of phenomena not accounted for.

That being said, we now turn our attention towards an example of translation zone iteration that led to an update at the level of the candidate model construction. To understand that, we must first note the fact that the device model presented in the above subsection did not always have this configuration. Originally, a much higher focus has been placed on modelling the internal hardware components of devices, in an attempt to construct a representation that would explicitly provide more details about the underlying recovery methods used. From the perspective of the translation zone, we can argue that the first iteration led to a conceptual candidate model representation in which each device was seen as a separate model, with distinct memory areas as locations, various software operations such as initializing the BIOS or verifying a security certificate as processes, and different software resources such as encryption keys, log files and so on. However, an analysis of this representation has shown that this high level of detail would not produce significant behaviour for the overall organisational recovery target, because an extended set of timing measurements for this highly specialized hardware process were unavailable, and very hard to measure --- some of the operations were taking place before the operation system started. This was accounted for in the scope alteration phase of the translation zone, and the new model scope was updated to reflect the new conceptualization for the device model, as presented above. In turn, this resulted in a re-implementation of the device model in the next cycle iteration.

The resulting simulation model was executed over 9000 different parameter configurations, totalling an amount of 450000 iterations. In the context of this paper, we describe one of the model consequences resulting from the model execution: under the exponential category of attacks with a high infection probability, allocating two admins to help-desks in large offices produced better recovery results than allocating three admins. This seemed contradictory, given the fact that admins are directly involved in the recovery procedures, and waiting for an admin to become available can be one of the reasons for high recovery timings. Therefore, the verification process at the level of the model produced a contradiction, but the analysis of input data and code did not reveal any issues. During domain consequences translation, this was also discussed by the modellers and research manager, then validated with the help of the security architect and help-desk specialist at the level of the domain. Direct analysis of the results showed something interesting: in the case of three admins being present in a large office, there were moments when two of them were being deployed to smaller locations to help with recovery and only a single one remained. However, when only two were present, the workload was high enough so they were not able to move. Discussions with the help-desk specialist revealed the absence of a policy for this specific case, which in turn indicated that the situation was possible in practice, yet unaccounted for. 

Therefore, the analysis of outputs produced by a model constructed using the distributed systems metaphor and co-design approach --- namely the organisational recovery model --- was used to determine a real possible area of improvement which might be relevant for any organisation that lacks a detailed admin deployment policy under ransomware attacks. Furthermore, the model produced outputs that can be interpreted similarly to the Sophos 2024 State of Ransomware report~\citep{sophos2024}. Although the report does not describe the nature and severity of encountered ransomware, it states that 35\% of surveyed organisations took between 1 and 6 months to recover. In the ransomware recovery model, the average recovery duration across all scenarios, but at a fixed 50\% infection probability is 69.4 days, which translates to about 3 months. However, since the Sophos report surveys organisations between 100 and 5000 employees and the model represents organisations with 120 employees, we can argue that the 35\% rate is higher when considering only smaller size organisations with less resources.

\subsection{Trauma unit surge capacity model}
    
The third model we introduce was used to explore the surge capacity of a trauma unit within a generic hospital emergency department. In comparison to the other two models described above, the trauma unit surge capacity model can be seen as simpler at the level of structure, but relevant for illustrating how the domain exploration and translation towards a candidate model can be performed in a context in which the modellers do not possess a high amount of experience or expertise. 

A trauma unit stands apart from regular medical services by specializing in the immediate and critical care of patients who have sustained severe injuries, typically due to accidents, violence, or other traumatic events. These units are specifically equipped and staffed to handle complex and life-threatening cases, often involving multiple injuries affecting different parts of the body. However, this is exactly why under crisis conditions, trauma units may become flooded with an influx of patients that can overwhelm their operational capacity. 

For instance, after a major incident, such as a train crash or a terrorist attack, a large number of critically injured patients may arrive at the hospital in a short amount of time. Surge capacity refers to the number of patients that can be treated before the quality of care declines to unacceptable levels. The model is used to explore how different factors, such as staffing levels, staff skills and experience, or available equipment, affect surge capacity and provide insight into how capacity can be increased. 

\begin{figure}[hbt!]
    \centering
    \includegraphics[scale=0.6]{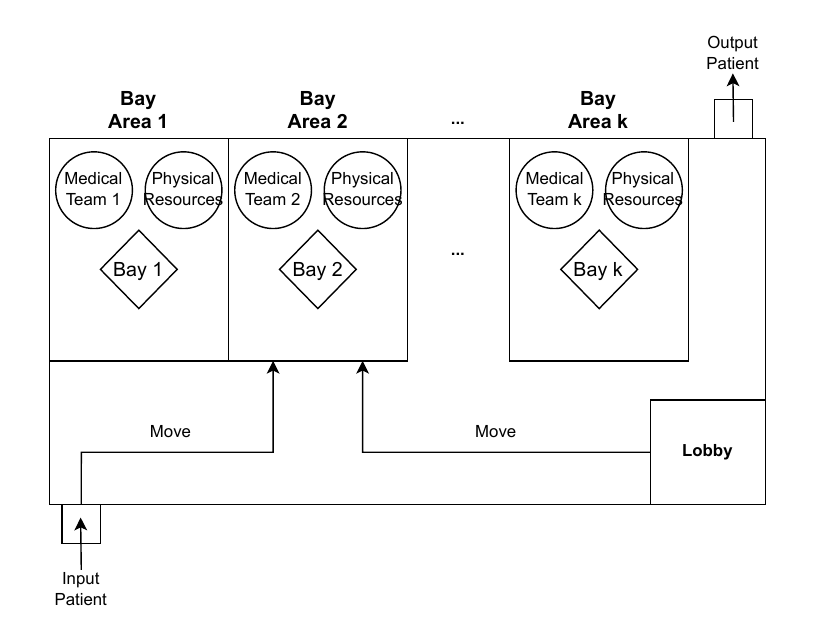}
    \caption{The trauma unit surge capacity model}
    \captionsetup{justification=centering}
    \label{fig:trauma_mod}
\end{figure}
\FloatBarrier 

\subsubsection{Structure \& representation choices}

As previously stated, structurally, the model is relatively simple, in the sense that no sub-model compositions are being performed. This can be easily observed in Figure~\ref{fig:trauma_mod}, where the model representation focuses on the physical and operational characteristics of a trauma room, but without conceptualizing other hospital areas that might be connected to it, such as perhaps a triage area, or the medical bays where the patients might be transferred to after no longer requiring emergency care.

In the model, patients, staff, equipment, and supplies are viewed as resources. Depending on a patient's injuries, different treatments, modelled as processes, need to be provided. These treatment processes require various resources, such as staff with particular skills, in order to execute. These resources are the limiting factor for surge capacity. For example, when there is a shortage of staff with a particular skill patients' treatment might be delayed, leading to undesirable outcomes.

In terms of \emph{locations}, the structure for this model is quite straightforward.  There are locations for each treatment bay area, a lobby where medical personnel are awaiting patients, the emergency department, its entrance, and one representing the rest of the hospital, where patients move after treatment.

Similarly, \emph{resources} are intuitively simple. There are resources representing patients, hospital staff, and available treatment bays. Each patient resource has an associated set of procedures that are required for treatment; each staff resource has a set of skills --- with an associated skill level --- which define their capabilities in terms of treatment.

With respect to \emph{environments}, they are again being used for starting the patient resource arrival process according to different probability distributions.

\emph{Processes} are slightly more complicated. Each patient has an associated process that moves the patient resource from outside the hospital, into the trauma room, and into a treatment bay when one is available.

Another process assigns the teams to the treatment bay.  This process encodes the decision-making around team formation as learned from the hospital staff during model construction.

When the staff resources are moved to a treatment bay area, treatment can begin. As mentioned above, each patient resource has a set of procedures that describe their treatment. Each of these procedures is modelled using a different process. Each of these processes requires staff resources with different skills. For example, the team leader process requires someone with a high level of the `team leader' skill --- experience in leading trauma teams; or, the intubation procedure (called `advanced airway' in the model) requires someone with the `advanced airway' skill at a suitable level.  The procedures have dependencies.  For example, the `advanced airway' procedure can only begin after the `patient assessment', `airway assessment', and `IV' procedures have completed.

Procedures can execute in parallel as long as their dependencies have been met and sufficient staff resources with appropriate skills are present. When team sizes are reduced because of higher patient numbers, some procedures might have to wait until a staff member becomes free, which extends the duration of the patient's treatment.

This representation of procedures, although complex, was chosen as it allows the effects of staff limitations to be captured by the model.

\subsubsection{Co-design}

After a brief examination of the model's structure and associated representation choices, we redirect our focus towards illustrating some of the co-design elements at play. Given the less familiar nature of the topic area for modellers, we describe two approaches employed to achieve a better understanding both during domain exploration and within the translation zone: a practice dummy exercise and a tabletop exercise.

From the very beginning, we can clearly state that co-design featured prominently in this work, even at goal level. For example, the initial question about surge capacity was brought to the modelling team --- who at that time knew nothing about hospital operation and management --- by a consultant anaesthetist from the hospital, who then worked closely with the modelling team throughout the process. Additionally, other medical personnel were involved in the process as part of the modelling team, on a case-by-case basis, but mediated with the help of the consultant anaesthetist. 

As part of the domain exploration phase of the co-design cycle, the modelling team visited the hospital on many occasions, observing the operation of the trauma unit as well as other aspects of hospital operation. To better understand the decision-making process behind the formation of treatment teams, and how various treatment procedures are performed on patients, the modelling team observed a simulated treatment of a trauma patient on a training dummy --- which is standard practice in hospitals to maintain skills. This revealed an important aspect regarding the treatment processes: different treatment procedures are temporally subordinated to other treatment procedures. In other words, for certain treatment procedures to be performed, some other treatment procedures must be first completed --- to perform the `advanced airway' procedure on a patient requires the `patient assessment', `airway assessment', and `IV' procedures to be completed first.

In terms of data, the primary source for the modelling team was the consultant anaesthetist working closely with them. This was further supplemented or verified by discussion with more hospital staff members, and ranged from informal discussions with hospital colleagues about various aspects of the model --- duration of procedures, accuracy of the flow of patients as described by the models --- to a more formal tabletop exercise. 

The tabletop exercise was designed to gain insight into how teams of staff are formed to treat patients, and how team formation and re-allocation changes when additional patients arrive. The exercise involved sticker notes representing members of staff and patients. We would begin the exercise with a single patient and asking the subject to form a team using the cards. We then introduced additional patients and asked how they would form new teams, and whether they would remove staff members from existing teams. From a co-design perspective, this can be understood as an attempt to infer domain behaviour from an artificial version of the domain. Nonetheless, this did not represent an issue, because the present medical staff explained the possible limitations of this environment, and how conditions could change during real treatment. 

With respect to translation, it is important to note that the consultant anaesthetist learned how to express ideas in terms of the distributed system metaphor --- for example, by thinking in terms of the process of patient treatment, and the resources (staff, skills, equipment) required for different procedures. Furthermore, he ensured the communication between modellers and other medical staff led to a translation of concepts towards the metaphor, without relevant information being lost. 

In this context, the most relevant quantitative data used in the simulations was related to the medical procedure timings. In the current version of the model, the timings are quantified using ranges estimated by trauma unit staff. Future work will employ a nurse or other staff member to record accurate timings for different procedures to be used in the model. To ensure the validation procedure is understood and can be performed with the help of medical staff, a graphical visualisation of the model was produced, showing patients moving into treatment bays, the formation of teams, and the progress of the various treatment procedures. This was then shown to hospital staff to get feedback and to validate the behavior of the model. We note here that the model did not attempt to construct a generalized representation of any trauma room, but rather focused on producing targeted insights regarding the specific procedures observed in the specific trauma room available. Therefore, during validation, the medical staff working in that trauma room assessed whether or not the model was producing `believable' outputs, similar to those they encountered on a daily basis. 

Lastly, after the model was constructed, it was used to simulate the trauma unit under real-world conditions as exhibited during the 2017 London Bridge terror attack --- this included data regarding patient load, injury type, and arrival rate. In this setting, the model produced similar behaviour to the actual trauma room --- validated by staff --- and even produced some insights about the expected difference in surge capacity for daytime and nighttime staffing levels.

\subsection{Reflections}

Previously, we have attempted to illustrate how the distributed systems conceptualization and co-design methodology were used to practically design and construct the three presented models. Here, we draw considerations by reasoning about how the concepts of the metaphor and co-design steps were employed in the models.

We start by analysing how the properties of the metaphor, as shown in Section~\ref{subsec:ds_met_prop} manifested at the level of the models.

In terms of \emph{generality}, it can be claimed without a doubt that the distributed systems concepts were suitable for constructing working representations of the systems under study. Irrespective of the characteristics of the target system, whether it was abstract and technical as seen in the ransomware recovery model or more centered on human factors as observed in the trauma unit surge capacity model, the metaphor successfully encapsulated all pertinent entities and relationships according to the model goal. Furthermore, a high degree of flexibility can be observed: for instance, in the data-loss model, resources are mostly physical (laptops, cds, USBs), whereas in the recovery one, they can be physical (devices, USBs), abstract (recovery images, network data), or even people (help-desk employees). Additionally, hierarchies of relationships can be constructed between concepts of the same type. In the recovery model, both recovery images and recovery agents resources are encoded in network data resources to be then routed through the network model. Similarly, in the trauma unit model, starting the various treatment processes may depend on the completion of other treatment processes beforehand. Finally, as observed in the data-loss model, the concepts themselves can be bundled to produce new abstractions, such as the one for agents, involving a resource, a process and multiple abstract locations. 

With respect to \emph{recognizability}, the interaction between modellers, stakeholders and practitioners in both the ransomware recovery and trauma unit models has shown that the elements of the distributed systems metaphor are relatively easy to grasp, especially since they are general, and versions of their definitions already exist in the scientific areas that the stakeholders and practitioners were familiar with. For example, at the level of the recovery model, the research manager did not have any issues in conceptualizing the physical and abstract resources immediately, but a longer discussion was needed for help-desk employees. 

Similarly, we have observed that the \emph{identity conservation} criterion was not only satisfied, but that the constructed identity for the model components, based on the distributed systems metaphor, facilitated focused and clear discussions between modelling participants during the entire co-design process. This was showcased this at the level of the recovery model, where originally devices were conceptualized as models with a complex internal structure, then simply as resources. Although the original representation was more detailed, that level of detail did not provide relevant information from an organisational recovery perspective. Furthermore, because the available information sources did not explicitly specify the precise effects of various ransomware strains at the level of individual hardware components, the first representation ended up distracting the modellers from the most relevant recovery aspects: the time of device infection and the recovery duration. However, once the device conceptualization was changed in a subsequent cycle iteration, and the modelling team members became familiar with the new device identity, the rest of the translation zone processes became simpler, because they did not require integration with the removed aspects of the device representation. 

Regarding the \emph{scale-freeness} property, we can argue that it was manifested in a set of different ways. Firstly, at the level of the concepts: both resources and locations were treated similarly, regardless of their physical or abstract nature or real-world scale. Secondly, all the models' scale can be adjusted with the help of model parameters. For example, in the trauma unit model, the number of available medical personnel, the number of treatment bays or the patient arrival rate can be set through external parameters. This can lead to simulations representing very small trauma units, or large trauma departments by essentially using the same structure. Lastly, \emph{scale-freeness} is also supported via the formal composition operation. For instance, we can easily envision the emergency trauma unit model as being composed with a triage model on the patient input side, and with other hospital departments on the output side, in an attempt to explicitly represent an entire hospital. 

Turning our attention to co-design aspects, we further describe some of the observed benefits, but also difficulties encountered.

Firstly, in both the cases of the ransomware recovery and trauma unit models, the object of inquiry was not decided by modellers alone, but rather was provided by organisational stakeholders like the security research manager and consultant anaesthetist, and then aligned with the research goals of the modellers during the construction of the model scope. Because of that, the stakeholders had a direct interest in the success of the model development, and helped the modelling team not only with domain knowledge, but also by facilitating meetings with other experts and stakeholders. This explicitly improved the domain exploration and validation steps, by allowing the modellers to have access to more information sources and providing experience-based opinions where necessary. Also, it mitigated one of the primary difficulties encountered, namely the impossibility of all the stakeholders to be continuously present in the modelling process. Although this was not the case in the showcased model examples, our belief is that the overall model quality might have been reduced if these stakeholders would not have been interested and actively engaged in the modelling task. 

Secondly, both the ransomware recovery and trauma unit models have produced representations, behaviours, and results comparable to their real-world targets. In both the case of ransomware recovery and trauma unit models, this has been shown via comparisons with the Sophos 2024 ransomware report~\citep{sophos2024}, and with the recorded data about the exhibited behaviour in the trauma unit during the 2017 London Bridge terror attack, as described in the above subsection. Additionally, the results produced by the ransomware recovery model were used to determine a possible admin policy issue that was not even originally considered or the main goal of the model. Therefore, this showcases the ability of the modelling methodology to produce models that can facilitate the discovery of phenomena outside of the explicit model scope, but with a causal influence on it. The absence of a limitation in admin behaviour was noted precisely because it had an impact on the overall recovery duration, but since the stakeholders --- and specifically the help-desk specialist --- knew that such a policy did not exist, the model did not accounted for it either. 

Last but not least, it is important to acknowledge the occurrence of mutual learning throughout the co-design process. For example, during the development of the ransomware recovery model, the research manager installed, ran, and analysed model outputs --- with the help of the modellers --- in a programming language with which he was not familiar with, and the modellers learned about state-of-the-art hardware recovery mechanisms. Similarly, in the trauma unit case, the modellers learned about healthcare facility operation and treatment procedures, whereas the consultant anaesthetist got familiarized with the design, construction and interpretation of models under our presented approach.

\section{Conclusion}\label{sec:conc}

In this article, we have focused on illustrating a methodological approach to modelling heterogeneous systems inspired by a metaphor of distributed systems from computer science. The argument justifying this decision can be summarised as follows: 

If the objects of modelling continue increasing in complexity and heterogeneous systems maintain their importance for the average persons' life, then modelling approaches will have to adapt to these new targets as well. Given the heterogeneity aspect, this will most likely require integration between modelling traditions of various scientific disciplines. For that to be possible, the exploration of the domain in which the object of modelling is manifested and the relationship between model and goals must take into account notions that span across multiple disciplines. We argue that such candidates are the means of construction of both the observed phenomena and model, which can provide a neutral interpretation basis when underlined by an inferentialist perspective. Furthermore, the principles of co-design describe ways in which such integration can be achieved, but require the participants to have an interest in pluri-perspectivist understanding of the phenomena under study and to partake in knowledge sharing during the process. We cannot assume a priori that modelling participants posses these interests, but we can attempt to instill them during the modelling process. In an attempt to do so, we conceptualise the translation zone of our co-design cycle as a trading zone and then employ the concepts of the distributed systems metaphor as a cultural, in-between language tool --- this would not be possible if the trading zone and distributed systems metaphor would not be compatible --- to facilitate knowledge sharing and interactional expertise development and, therefore, driving the evolution of the trading zone from initially fractionated, towards inter-language. 

Following this argument, we consider the contributions of this research article to be significant.

Firstly, in Section~\ref{sec:background} we briefly analyse referentialist, inferentialist and more pragmatic, engineering-focused philosophic positions regarding the nature of models and determine that an inferentialist stance represents a good candidate for understanding and constructing heterogeneous models. 

Secondly, in Section~\ref{sec:what_model} we describe our notion of model and construct a qualitative metric for describing models grounded in inferentialism and specifically focused on the means of construction of both observed phenomena and constructed models.

Thirdly, in Section~\ref{sec:co_design} we identify caveats in the classical, mathematical modelling cycle and construct our own version of co-design cycle in an attempt to solve them for the case of heterogeneous models. This requires direct integration with the above metric and explicit inclusion of co-design principles at the level of the translation zone.

Fourthly, in Section~\ref{sec:ds_met} we attempt to increase the level of development of interactional expertise and knowledge sharing between modelling participants by structuring the translation zone with the help of the distributed systems metaphor. To do so, we compare the metaphors of trading zones and distributed systems, determine that they are compatible and then employ a strategy of guiding the evolution of the trading zone towards an inter-language one via an in-between language cultural tool, represented by the distributed systems conceptualisation. 

Lastly, as with any scientific hypothesis or methodology claiming to have real-world outcomes, abstract reasoning about its validity should not be the sole reason justification for its correctness, or even its utility. In Section~\ref{sec:case_studies}, we have showcased the practical applicability and effectiveness of our method by putting it to the test in three different environments that are strongly related to systems' security and resilience. The analysis of model results and the interaction with stakeholder during the modelling process have shown that the method is suitable for security modelling for a set of reasons: the constructed models helped the participants to understand better the underlying systems, were representationally aligned with the participants' perspectives about the systems, and, in the case of the ransomware recovery model, led to the discovery of a relevant, unaccounted for factor for recovery --- the admin deployment policy ---, and to similar overall results as the Sophos 2024 State of Ransomware report~\citep{sophos2024}, which was released after the model was constructed.

We note that the distributed systems metaphor is not only a well suited abstraction for a general notion of system, but also supports a series of much needed practical properties such as the ability to perform compositions, substitutions or local reasoning which have been formalised in prior work and are freely available to use practically as a Julia SysModels~\citep{juliacode} programming library.

Furthermore, we argue that our approach is suitable for heterogeneous system modelling more generally, for two primary reasons. On the theoretical side, co-designing models using the distributed systems metaphor satisfies the main criteria for the functioning of a trading zone as shown by~\citep{collins2007}: the underlying model components maintain a common identity notion across the modelling participants based on the distributed systems conceptualisation. On the practical side, the cases studies models introduced in Section~\ref{sec:case_studies} showcase the adaptability of our method in relatively complex and diverse representational settings. We are aware that exemplification alone cannot be seen as a complete proof for quality, but we believe that the theoretical argument, corroborated with the useful properties introduced and the positive results obtained up to this point should at least make our methodological approach to heterogeneous modelling a solid candidate for further experimentation. 

Nevertheless, much work still remains to be done. For example, the current version of the presented co-design cycle could be used to identify extended areas where bias can be introduced during model design and construction. Furthermore, the relationship between the different means of model construction, their underlying language components and strategies for the justification of belief in the spirit of inferentialism, across different scientific areas, should be further analysed. Additionally, the provided software modelling package could be further enhanced by an addition of dynamic model checking capabilities. Last but not least, more case studies should be conducted, with the particular focus on comparing models constructed using our approach with models constructed using different methods in an attempt to assess the quality of their outputs, over time, in competitive environments. 

\subsection*{Acknowledgements}

We are grateful to Adrian Baldwin, Kevin Fong, and Will Venters for their advice on various aspects of this paper.

\bibliography{bibliography}






\subsection*{Funding}

This work was partially supported by the UK EPSRC through a PhD Studentship and by research grant EP/R006865/1.

\subsection*{Competing Interests}

The authors have no relevant financial or non-financial interests to disclose.



\end{document}